\documentclass[aps,prd,groupaddress,floatfix,tighten,nofootinbib,showpacs,amsfonts,superscriptaddress]{revtex4}
\usepackage{url}
\usepackage{xcolor}
\usepackage{hyperref}

\colorlet{shadecolor}{gray!15}
\definecolor{greenLinks}{rgb}{0, 0.6, 0}
\definecolor{blueLinks}{rgb}{0, 0, 0.6}
\definecolor{redLinks}{rgb}{0.6, 0, 0}
\definecolor{tempText}{rgb}{0.55, 0.10,0.67}
\definecolor{eprintLinks}{rgb}{0.4, 0.4, 0.4}
\definecolor{journalLinks}{rgb}{0.6, 0, 0}
\usepackage{amssymb}
\usepackage{amsmath,color}
\usepackage{epsfig}
\usepackage{graphicx,epsf,epsfig}
\usepackage{footnote}
\usepackage{multirow}
\usepackage{enumitem}
\usepackage{color}
\def\slc#1{\setbox0=\hbox{$#1$}                  
    \dimen0=\wd0                                 
    \setbox1=\hbox{/} \dimen1=\wd1               
    \ifdim\dimen0>\dimen1                        
       \rlap{\hbox to \dimen0{\hfil/\hfil}}      
       #1                                        
    \else                                        
       \rlap{\hbox to \dimen1{\hfil$#1$\hfil}}   
       /                                         
    \fi}

\def\be{\begin{equation}}
\def\ee{\end{equation}}
\def\gs{\mathrel{
   \rlap{\raise 0.511ex \hbox{$>$}}{\lower 0.511ex \hbox{$\sim$}}}}
\def\ls{\mathrel{
   \rlap{\raise 0.511ex \hbox{$<$}}{\lower 0.511ex \hbox{$\sim$}}}}

\newcommand{\ba}{\begin{array}{c}}
\newcommand{\baz}{\begin{array}{cc}}
\newcommand{\barrr}{\begin{array}{rrr}}
\newcommand{\bad}{\begin{array}{ccc}}
\newcommand{\bav}{\begin{array}{cccc}}
\newcommand{\baf}{\begin{array}{ccccc}}
\newcommand{\bea}{\begin{equation} \begin{array}{c}}
\newcommand{\eea}{ \end{array} \end{equation}}
\newcommand{\ea}{\end{array}}

\usepackage[normalem]{ulem}

\def\21{$\mathrm{SU(2)_L \otimes U(1)_Y}$ }

\newcommand {\ignore}[1]{}
\newcommand{\lra}{\mu \leftrightarrow \tau}

\newcommand{\nn}{\nonumber}
\allowdisplaybreaks \allowdisplaybreaks[2]
\newcommand{\AddrIFUNAM}{
 Instituto~de F{\'{\i}}sica, 
 Universidad~Nacional Aut\'onoma de M\'exico, \\
 Apdo. Postal 20-364, CDMX 01000, M\'exico.}
  
\newcommand{\AddrFCEBUAP}{
 Fac. de Cs. de la Electr\'onica, 
 Benem\'erita Universidad Aut\'onoma de Puebla, 
 Apdo. Postal 542,\\ Puebla, Pue. 72000, M\'exico.}

\newcommand{\AddrTecMty}{
 Tecnologico de Monterrey, Campus Estado de 
 Mexico, Atizapan de Zaragoza, Estado de Mexico, Apartado Postal 52926, Mexico.
} 
\newcommand{\AddrCINVESTAV}{
Departamento de F\'isica, Centro de Investigaci\'on y de Estudios Avanzados del I. P. N.,\\
Apdo. Post. 14-740, 07000, Ciudad de M\'exico, M\'exico.}

\newcommand{\AddrCIFFU}{Centro Internacional de F\'{\i}sica Fudamental, 
 Benem\'erita Universidad Aut\'onoma de Puebla.
} 
  
\begin{document}
\title{On~${\mathbf Q}_{6}$ flavor symmetry and the breaking of $\lra$~symmetry} 
%
\author{Juan Carlos G\'omez-Izquierdo}
\email{jcgizquierdo1979@gmail.com}
\affiliation{\AddrTecMty}\affiliation{\AddrIFUNAM}
\affiliation{\AddrCINVESTAV}
\author{F. Gonzalez-Canales}
\email{felixfcoglz@gmail.com}
\affiliation{\AddrFCEBUAP}
\affiliation{\AddrCIFFU}
\author{M. Mondrag\'on}
\email{myriam@fisica.unam.mx}
\affiliation{\AddrIFUNAM}
%

\begin{abstract}\vspace{2cm}
 In the simplest version of a $\mathbf{Q}_{6}$ flavored supersymmetric model, we analyze 
 the leptonic masses and mixings in the framework of a soft breaking of the $\lra$ symmetry. This 
 breaking is controlled by the inequality $m_{e\tau}\neq m_{e\mu}$ in the effective neutrino 
 mass. As a consequence of this breaking, the reactor and 
 atmospheric angle are deviate from $0^{\circ}$ and $45^{\circ}$, respectively. Such deviations can be enhanced or suppressed by the CP parities in the Majorana phases, 
 so that an analytic study is carried out to remark their importance to constrain the free 
 parameters that accommodate the mixing angles. The normal
 hierarchy is completely discarded in this model, the inverted hierarchy is less favored  
 than the degenerate one where the reactor and atmospheric angles are in good agreement 
 with the experimental data.
Additionally, the model predicts defined regions for the effective neutrino mass decay, 
 the neutrino mass scale and the sum of the neutrino mass in the inverted and degenerate mass spectrum. Thus, this model may be testable by future  
 experiments that focus in neutrinoless double beta decay.
\end{abstract}

\begin{flushright}
CIFFU-17-04
\end{flushright}
%
\maketitle
%

\section{Introduction}
In the last two decades the neutrino oscillation experiments have provided a large amount of 
evidence in favor of the massive neutrinos and leptonic flavor mixings. 
In the theoretical framework of three active neutrinos we need only six independent parameters 
in order to characterize the so-called ordinary neutrino oscillations. These parameters are:
the difference of the squared neutrino masses, the flavor mixing angles and the ``Dirac-like'' 
CP violation phase factor~\cite{Giunti:Book}. 
The numerical values for the squared neutrino masses and flavor mixing angles obtained from a 
global fit to the current experimental data on neutrino oscillations, at Best Fit Point (BFP) 
$\pm 1 \sigma$ and $3 \sigma$ ranges, are~\cite{Forero:2014bxa}:
\begin{equation}
 \begin{array}{ll}\vspace{3mm} 
  \Delta m^{2}_{21} \left( 10^{-5} \, \textrm{eV}^{2} \right) = 
   7.60_{-0.18}^{+0.19}, \, 7.11 - 8.18, &
  \left| \Delta m^{2}_{31} \right| \left( 10^{-3} \, \textrm{eV}^{2} \right) =  
   \left\{ \begin{array}{l}\vspace{2mm}
    2.48_{-0.07}^{+0.05}, 2.30 - 2.65  \\ 
    2.38_{-0.06}^{+0.05}, 2.20 - 2.54
   \end{array} \right. , \\ \vspace{3mm}
  \sin^{2} \theta_{12} / 10^{-1} = 3.23 \pm 0.16, \, 2.78 - 3.75 , &
  \sin^{2} \theta_{23} / 10^{-1} = 
   \left\{ \begin{array}{l}\vspace{2mm}
    5.67_{-1.24}^{+0.32} , \, 3.93 - 6.43 \\ 
    5.73_{-0.39}^{+0.25} , \, 4.03 - 6.40
   \end{array} \right.  , \\ \vspace{2mm}
  \sin^{2} \theta_{13} / 10^{-2}  = 
   \left\{ \begin{array}{l}\vspace{2mm} 
	2.26 \pm 0.12 , \, 1.90 - 2.62  \\  
	2.29 \pm 0.12 , \, 1.93 - 2.65
   \end{array}  \right. . 
 \end{array}
\end{equation}
The upper and lower rows are for a normal and inverted hierarchy of 
the neutrino mass spectrum, respectively. Currently, the experimental determination of the 
above parameters is in a precision age. However, this is not the case for the ``Dirac-like'' CP 
violation phase in the leptonic sector, since the T2K experiment only give us hints of an approximately 
maximal CP violation phase, $\delta_{\mathrm{CP}} \sim -\pi/2$~\cite{Abe:2013hdq}. 
Then, the $\delta_{\mathrm{CP}}$ phase from a global fit to the current experimental data on 
neutrino oscillations, at BFP$\pm 1 \sigma$ and $3 \sigma$ level, is~\cite{Forero:2014bxa}:
\begin{equation}
 \begin{array}{ll}\vspace{3mm} 
  \delta_{\mathrm{CP}} / ^{\circ}  = 
   \left\{ \begin{array}{l}\vspace{2mm}
    254_{-72}^{+99} , \, 0 - 360 \\ 
    266 \pm 56 , \, 0 - 360 
   \end{array}  \right.   . 
 \end{array}
\end{equation}

The above experimental evidence was enough to show that 
neutrinos have a tiny mass, whereby it was very easy to conclude that there is 
physics beyond the Standard Model~(SM). Now we have the dilemma how far beyond the SM we 
have to go and under what arguments this extension is based. In this work we consider 
the current experimental data on neutrino oscillations that were mentioned above. So, 
from these data  we obtain that the magnitude of the leptonic mixing matrix elements have 
the following values at $3\sigma$
\begin{itemize}
 \item Normal Hierarchy (NH)
  \begin{equation}\label{Eq:PMNS:fits-1}
   \left( \begin{array}{ccc}
    0.780 - 0.842 & 0.520 - 0.607 & 0.137 - 0.162 \\
    0.207 - 0.555 & 0.395 - 0.714 & 0.618 - 0.794 \\
    0.226 - 0.566 & 0.420 - 0.731 & 0.590 - 0.772 
   \end{array}  \right).
  \end{equation}
 \item Inverted Hierarchy (IH)
  \begin{equation}\label{Eq:PMNS:fits-2}
   \left( \begin{array}{ccc}
    0.779 - 0.842 & 0.520 - 0.607 & 0.139 - 0.163 \\
    0.207 - 0.554 & 0.397 - 0.710 & 0.626 - 0.792 \\
    0.229 - 0.566 & 0.426 - 0.729 & 0.592 - 0.765 
   \end{array}  \right) .
  \end{equation} 
\end{itemize}
From the leptonic mixing matrices given in the Eqs.~(\ref{Eq:PMNS:fits-1}) 
and~(\ref{Eq:PMNS:fits-2}) we obtain for the second and third row the 
remarkable result
\begin{equation}
 \left| U_{\mu 1} \right| \approx \left| U_{\tau 1} \right|, \qquad
 \left| U_{\mu 2} \right| \approx \left| U_{\tau 2} \right|, \qquad
 \left| U_{\mu 3} \right| \approx \left| U_{\tau 3} \right|. 
\end{equation}
This means that the leptonic mixing matrix has in an approximate way the so called 
$\lra$ symmetry.
In the model building context, the $\lra$ flavor symmetry has been widely used to 
propose possible extensions of the SM. In these extensions, the $\lra$~flavor 
symmetry can be defined in two different ways: 
\begin{itemize}
 \item[$i)$] the $\lra$ {\it permutation symmetry}~\cite{Mohapatra:1998ka, Lam:2001fb, Kitabayashi:2002jd, Grimus:2003kq, Koide:2003rx,Fukuyama:1997ky} where the 
  neutrino mass term is unchanged under the transformations 
  $\nu_{e} \rightarrow \nu_{e}$, $\nu_{\mu} \rightarrow \nu_{\tau}$ and
  $\nu_{\tau} \rightarrow \nu_{\mu}$. 
 \item[$ii)$] the $\lra$ {\it reflection symmetry}~\cite{He:2015xha, Xing:2015fdg, Nishi:2016wki} where the 
  neutrino mass term is unchanged under the transformations 
  $\nu_{e} \rightarrow \nu_{e}^{c}$, $\nu_{\mu} \rightarrow \nu_{\tau}^{c}$ and
  $\nu_{\tau} \rightarrow \nu_{\mu}^{c}$, where $c$ denotes the charge conjugation. 
\end{itemize}
In here we will only consider the first definition, hence in the following when we mention the $\lra$ symmetry actually we mean  the $\lra$ 
permutation symmetry.
Historically,  theoretical physicists have proposed the $\lra$ symmetry  in 
order to reproduce the experimental data on lepton mixing angles. Namely, the $\lra$ 
symmetry is obtained if the neutrino oscillation parameters fulfill one of the 
following conditions 
\begin{equation}
 \left| U_{\mu i} \right| = \left| U_{\tau i} \right|
 \quad \Longleftrightarrow \quad
 \left \{
  \begin{array}{lcl}
   \theta_{13} = 0^{\circ} & \textrm{and} & \theta_{23} = 45^{\circ}, \\
   \delta = \pm 90^{\circ} & \textrm{and} & \theta_{23} = 45^{\circ}.
  \end{array} \right.
\end{equation}
The first one is ruled out by the current experimental data on neutrino oscillations, but not 
even the second condition is allowed at the $1\sigma$ or $3\sigma$ level. 
At present, the Long-baseline energy experiment NO$\nu$A has disfavored the exact $\mu-\tau$ 
symmetry, therefore some possible breakings of the $\lra$ symmetry have 
been explored~\cite{Haba:2006hc, GomezIzquierdo:2007vn, Dicus:2010yu, Ge:2011ih, Ge:2011qn, He:2011kn, Araki:2011zg, He:2012yt, Hanlon:2013ska, Ge:2014mpa, Luo:2014upa, Rivera-Agudelo:2015vza} (for generic models see~\cite{Xing:2006xa, Ge:2010js, Grimus:2012hu, Xing:2010ez, Gupta:2013it, Xing:2015fdg, Zhao:2016orh, Chen:2015siy, Chen:2016ica}). 
If this symmetry is broken, it is vital to investigate the source of the breaking and the 
framework where this is realized. \\

Based on the above, we build a supersymmetric model with the $\mathbf{Q}_{6}$ symmetry group as flavor 
symmetry and conserved $R-$parity. In this context, as a first step we study the masses and flavor mixing in the leptonic sector, 
where the $\lra$ symmetry is only broken in the effective neutrino mass matrix, by the 
difference $m_{e\tau}-m_{e\mu}\neq0$, which deviates the atmospheric angle from $45^{\circ}$ 
and the reactor angle is non zero.
In previous works on 
$\mathbf{Q}_{6}$~\cite{Babu:2004tn, Kajiyama:2005rk, Kajiyama:2007pr, Kifune:2007fj, Babu:2009nn, Kawashima:2009jv, Babu:2011mv}, the scalar sector was extended such that three families 
of doublets $H^{d}_{i}$ and $H^{u}_{i}$ are needed for the mixing. Contrary other models reported in the literature, we will try to explain the contrast between the CKM and PMNS mixing matrices by assigning in a different way the quark and lepton sector under the action of the flavor symmetry, as we will see below. The quark sector will be analyzed in a future work.
\section{The Model}
In our model the matter content and how it transforms under the action of the $\mathbf{Q}_{6}$ flavor symmetry is shown in Table~\ref{tabla1}. As we can see, the quark and lepton sectors have been assigned in a different way under the action of the flavor symmetry; the main reason to do so is to take seriously the remarkable hierarchy in the quark masses. Although the charged lepton masses may preserve this hierarchy, in the neutrinos sector this is not clear.
\begin{table}[!htbp]
 \begin{center}
 \begin{tabular}{|c|c|c|c|c|c|c|} \hline \hline 
  {$\bf Q_{6}$}  & {\bf $1_{+,0}$} & {\bf $1_{+,2}$} & {\bf $1_{-,1}$} & 
   {\bf $1_{-,3}$} & {\bf $2_{2}$} & {\bf $2_{1}$} \\ \hline 
  Matter & {\footnotesize $H^{d}_{3}$ } & 
   {\footnotesize $H^{u}_{3}$ }, {\footnotesize $Y_{B}$ } & 
   {\footnotesize $L_{1}$ }, {\footnotesize $N^{c}_{1}$ }, 
   {\footnotesize $Q_{3}$ }, {\footnotesize $u^{c}_{3}$ } & 
   {\footnotesize $\ell^{c}_{1}$}, {\footnotesize $d^{c}_{3}$ }  & 
   {\footnotesize $L_{J}$}, {\footnotesize $\ell^{c}_{J}$ }, 
   {\footnotesize $N^{c}_{J}$ }, {\footnotesize $Q_{I}$ }, 
   {\footnotesize $d^{c}_{I}$ }, {\footnotesize $u^{c}_{I}$ }, 
   {\footnotesize $H^{d}_{I}$ } & {\footnotesize $H^{u}_{I}$ } \\ \hline\hline 
 \end{tabular}\caption{Matter content.\label{tabla1}} 
 \end{center}
\end{table}
So the quark and Higgs superfields 
$Q_{I}$ and $H_{I}^{u,d}$, with $I=1,2$, transform as doublets. 
In fact, for the rest of the superfields, if they have the subscript $I$, it means that they transform as doublets under the flavor group, otherwise, the superfields transform 
as singlets under the flavor group. This flavor structure provides hierarchical quark mass matrices that reproduce the CKM mixing matrix quite well~\cite{Babu:2004tn, Kajiyama:2005rk, Kajiyama:2007pr, Kifune:2007fj, Babu:2009nn, Kawashima:2009jv, Babu:2011mv}. On the other hand, in the leptonic sector, the first family is assigned to any of singlets of $\mathbf{Q}_{6}$, while 
the other two families are assigned as elements of the flavor doublet, $L_{J}$ with $J=2,3$. In here, we would like to emphasize that such assignment is a possible route for realizing the $\lra$ symmetry \cite{ Grimus:2003kq, Grimus:2006jz, Adulpravitchai:2008yp, Hagedorn:2010mq, Araki:2011zg}, 
and the tribimaximal mixing matrix \cite{Harrison:2003aw, Mohapatra:2006pu}. In comparison to previous works on $\mathbf{Q}_{6}$~\cite{Babu:2004tn, Kajiyama:2005rk, Kajiyama:2007pr, Kifune:2007fj, Babu:2009nn, Kawashima:2009jv, Babu:2011mv}, where the assignment ${\bf 2}\oplus {\bf 1}$ (the first two families in the ${\bf 2}$ and the third one in the singlet irreducible representations) for the families of fermions and scalars was used, in this paper, we assigned differently the leptons (${\bf 1}\oplus {\bf 2}$, the first family in the singlet and the second and third one in the doublet) in order to identify and subsequently break the $\lra$ symmetry, as we will see later.

In this theoretical framework the superpotential has the form:
%
%
\begin{equation}\label{Yukb}
 \begin{array}{ll}
 {\bf W} =& 
   y^{u}_{1} \left( Q_{1} u^{c}_{2} 
   - Q_{2} u^{c}_{1} \right) H^{u}_{3} 
   + y^{u}_{2} \left( Q_{1} H^{u}_{2} 
   + Q_{2} H^{u}_{1} \right) u^{c}_{3}
   + y^{u}_{3} Q_{3} \left( u^{c}_{1} H^{u}_{2} 
   + u^{c}_{2} H^{u}_{1} \right)
   + y^{u}_{4} Q_{3} u^{c}_{3} H^{u}_{3} \\
  &
  + y^{d}_{1} \left[ Q_{1} 
   \left( - d^{c}_{1} H^{d}_{1} + d^{c}_{2} H^{d}_{2} \right) 
   + Q_{2} \left( d^{c}_{1} H^{d}_{2} + d^{c}_{2} H^{d}_{1} \right) \right] 
   + y^{d}_{2} \left( Q_{1} d^{c}_{1} + Q_{2} d^{c}_{2} \right) H^{d}_{3} 
   + y^{d}_{3} Q_{3} d^{c}_{3} H^{d}_{3} \\ 
  &
  + y^{\ell}_{1} L_{1} e^{c}_{1} H^{d}_{3} 
   + 	y^{\ell}_{2} \left[ L_{2} 
   \left( -e^{c}_{2} H^{d}_{1} + e^{c}_{3} H^{d}_{2} \right) 
   + L_{3} \left( e^{c}_{2} H^{d}_{2} + e^{c}_{3} H^{d}_{1} \right) \right]
   + y^{\ell}_{3} \left( L_{2} e^{c}_{2} + L_{3} e^{c}_{3} \right) H^{d}_{3} 
   + y^{D}_{1} L_{1} N^{c}_{1} H^{u}_{3} \\ 
  &
  + y^{D}_{2} L_{1} \left( N^{c}_{2} H^{u}_{2} + N^{c}_{3} H^{u}_{1} \right) 
   + y^{D}_{3} \left( L_{2} H^{u}_{2} + L_{3} H^{u}_{1} \right) N^{c}_{1}
   + y^{D}_{4} \left( L_{2} N^{c}_{3} - L_{3} N^{c}_{2} \right) H^{u}_{3}
   + y^{m} Y_{B} N^{c}_{1} N^{c}_{1} \\ 
  &
  + M_{R_{2}} \left( N^{c}_{2} N^{c}_{2} + N^{c}_{3} N^{c}_{3} \right)   
 \end{array} 
\end{equation}
Before going ahead we will make some important remarks. $i$)~In order to generate the flavor 
invariant Majorana mass term for the right-handed neutrinos, we included one Babu-Kubo flavon 
$Y_{BK}$~\cite{Babu:2004tn}. 
$ii$)~From the matter content given in Table I we obtain that the $\mu$-terms in the Higgs 
sector are not flavor invariant, whereby these kind of terms do not appear in the Higgs sector. 
However, the $\mu$-terms must be present in the Higgs sector because these are essential to get 
the electroweak symmetry breaking. Consequently, extra gauge singlets should be included in 
order to build the $\mu$-terms invariant under the action of the flavor group 
(see~\cite{LopezFogliani:2005yw, Fidalgo:2009dm}).
$iii$)~As the main aim of this work is the implication of the $\lra$ symmetry breaking in the leptonic masses and mixings, in this version of the model we assume a particular alignment in the vacuum expectation values (vev's) of the scalar fields.
%
\section{The lepton masses and flavor mixing angles}
%
From the superpotential given in Eq.~(\ref{Yukb}) the form of the Dirac fermions mass 
matrices in the leptonic sector is the following  
\begin{align}\label{ME}
 {\bf M}_{D} = 
 \begin{pmatrix}
  y^{D}_{1} \langle \textbf{H}^{u}_{3} \rangle & 
   y^{D}_{2} \langle \textbf{H}^{u}_{2} \rangle & 
   y^{D}_{2} \langle \textbf{H}^{u}_{1} \rangle \\ 
  y^{D}_{3} \langle \textbf{H}^{u}_{2} \rangle & 0 & 
   y^{D}_{4} \langle \textbf{H}^{u}_{3} \rangle \\ 
  y^{D}_{3} \langle \textbf{H}^{u}_{1} \rangle & 
   - y^{D}_{4} \langle \textbf{H}^{u}_{3} \rangle & 0
 \end{pmatrix}
 \quad \textrm{and} \quad
 {\bf M}_{\ell} = 
 \begin{pmatrix}
  y^{\ell}_{1} \langle \textbf{H}^{d}_{3} \rangle & 0 & 0 \\
  0 & y^{\ell}_{3} \langle \textbf{H}^{d}_{3} \rangle 
   - y^{\ell}_{2} \langle \textbf{H}^{d}_{1} \rangle & 
   y^{\ell}_{2} \langle \textbf{H}^{d}_{2} \rangle \\ 
  0 & y^{\ell}_{2} \langle \textbf{H}^{d}_{2} \rangle & 
   y^{\ell}_{3} \langle \textbf{H}^{d}_{3} \rangle 
   + y^{\ell}_{2} \langle \textbf{H}^{d}_{1} \rangle 
 \end{pmatrix}.
\end{align}
Here, ${\bf M}_{D}$ is the Dirac neutrinos mass matrix, while ${\bf M}_{\ell}$ is the 
charged leptons mass matrix.
In the flavor space, the Majorana right-handed (RHD) neutrino mass matrix has 
the diagonal form
\begin{equation}\label{Eq:MR}
 {\bf M}_{R} = 
  \textrm{diag} \left( M_{ R_{1} }, \, M_{ R_{2} }, \, M_{ R_{2} } \, \right)\, ,
\end{equation}
where $M_{ R_{1} } = y^{n} \langle Y_{BK} \rangle$.

As it is well known, the $\lra$ symmetry in the effective neutrinos mass matrix is identified very well in the basis where the charged lepton mass matrix is diagonal. In our model, this can be accomplished by the following alignment in the scalar sector
$\langle \textbf{H}^{u}_{1} \rangle = \langle \textbf{H}^{u}_{2} \rangle$, 
and $\langle \textbf{H}^{d}_{2} \rangle = 0$; at the same time such alignment allows us to reduce free parameters in the Dirac neutrino mass matrix,  and as a consequence in the effective neutrino mass matrix as well, where a partial $\lra$ symmetry is expected. Then, the mass matrices given in Eq.~(\ref{ME}) acquire the following forms, 
respectively,
\begin{align}\label{ME2}
 {\bf M}_{D} = 
 \begin{pmatrix}
  a_{D} & b_{D} & b_{D} \\ 
  c_{D} & 0 & d_{D} \\ 
  c_{D} & -d_{D} & 0
 \end{pmatrix}
 \quad \textrm{and} \quad 
 {\bf M}_{\ell} = 
 \textrm{diag} \left( a_{\ell}, b_{\ell},  d_{\ell} \right),
\end{align}
where 
$a_{D} \equiv  y^{D}_{1} \langle \textbf{H}^{u}_{3} \rangle $,
$b_{D} \equiv  y^{D}_{2} \langle \textbf{H}^{u}_{1} \rangle $,
$c_{D} \equiv  y^{D}_{3} \langle \textbf{H}^{u}_{1} \rangle $,
$d_{D} \equiv  y^{D}_{4} \langle \textbf{H}^{u}_{3} \rangle $,
$a_{\ell} \equiv y^{\ell}_{1} \langle \textbf{H}^{d}_{3} \rangle$, 
$b_{\ell} \equiv y^{\ell}_{3} \langle \textbf{H}^{d}_{3} \rangle 
   - y^{\ell}_{2} \langle \textbf{H}^{d}_{1} \rangle$,
and 
$d_{\ell} \equiv y^{\ell}_{3} \langle \textbf{H}^{d}_{3} \rangle 
   + y^{\ell}_{2} \langle \textbf{H}^{d}_{1} \rangle$.  
As the mass matrix ${\bf M}_{\ell}$ has a diagonal shape, the physical masses for the 
charged leptons are $m_{e} = \vert a_{\ell}\vert$, $m_{\mu} = \vert b_{\ell}\vert$ and 
$m_{\tau} = \vert d_{\ell}\vert$. So, all information about the leptonic flavor mixing 
only comes  from the neutrino sector.  

In this theoretical framework, the active Majorana neutrino mass matrix ${\bf M}_{\nu}$ 
is obtained through the type-I seesaw 
mechanism~\cite{GellMann:1980vs, Yanagida:1979as, Mohapatra:1980yp, Mohapatra:1979ia, Minkowski:1977sc, Schechter:1980gr, Schechter:1981cv}, 
${\bf M}_{\nu} = {\bf M}_{D} {\bf M}^{-1}_{R} {\bf M}^{\top}_{D}$, 
where the ${\bf M}_{D}$ and ${\bf M}_{R}$ matrices are given in Eqs.~(\ref{Eq:MR}) 
and~(\ref{ME2}). Hence, the explicit form of ${\bf M}_{\nu}$ is 
\begin{align}\label{mnu2}
 {\bf M}_{\nu} = 
  \begin{pmatrix}
   m_{ee} & m_{e\mu} & m_{e\tau} \\ 
   m_{e\mu} & m_{\mu\mu} & m_{\mu\tau} \\ 
   m_{e\tau} & m_{\mu\tau} & m_{\mu\mu}
  \end{pmatrix} \, , 
\end{align}
where 
\begin{equation}
 \begin{array}{lll}\vspace{2mm}
  m_{ee}     = 
   \frac{ a_{D}^{2} }{ M_{R_{1} } } + 2 \frac{ c_{D}^{2} }{ M_{R_{2}} } , &
  m_{e \mu}  = 
   \frac{ a_{D} b_{D} }{ M_{R_{1} } } - \frac{ c_{D} d_{D} }{ M_{R_{2}} } , & 
  m_{e \tau} = 
     \frac{ a_{D} b_{D} }{ M_{R_{1} } } + \frac{ c_{D} d_{D} }{ M_{R_{2}} } ,  \\
  m_{\mu \mu} =
   \frac{ b_{D}^{2} }{ M_{R_{1} } }  + \frac{ d_{D}^{2} }{ M_{R_{2} } } , &
  m_{\mu \tau}  =   \frac{ b_{D}^{2} }{ M_{R_{1} } } .
 \end{array}
\end{equation}
Remarkably, in this flavored model, the $\mu-\tau$ symmetry is only broken by the entries 
$m_{e \mu}$ and $m_{e \tau}$ ($m_{e\mu}\neq m_{e\tau}$). On the contrary, in generic 
models, where the charged lepton mass matrix is diagonal, the 
$m_{\tau\tau}\neq m_{\mu\mu}$ difference also breaks the $\mu-\tau$ symmetry; 
this implies extra free parameters in the effective neutrino mass matrix ${\bf M}_{\nu}$. 
Hence, this makes clear the advantage of working with the ${\bf Q}_{6}$ flavor symmetry. 
Let us go back to ${\bf M}_{\nu}$ matrix, the sub-block matrix $2-3$ provides a 
$45^{\circ}$ angle (to the flavor mixing matrix) that may be identified with the 
atmospheric one. 
Moreover, if the equality $m_{e \tau} = m_{e \mu}$ were true, the neutrino mass 
matrix, ${\bf M^{0}_{\nu}}$, would possess the $\mu-\tau$ symmetry. 
In addition, the ${\bf M^{0}_{\nu}}$ matrix would be diagonalized by means of the unitary 
transformation 
${\bf U}^{0 \dagger }_{\nu} {\bf M^{0}_{\nu}} {\bf U}^{0}_{\nu} = {\bf \Delta}_{\nu}^{0}$, where 
\begin{align}\label{e5e}
 {\bf U}^{0}_{\nu} = 
 \begin{pmatrix}
  \cos{\theta}_{\nu}  & 
  \sin{\theta}_{\nu}  & 0 \\ 
  - \frac{ \sin{\theta}_{\nu} }{ \sqrt{2} } & 
  \frac{ \cos{\theta}_{\nu} }{ \sqrt{2} } & 
  - \frac{1}{ \sqrt{2} } \\ 
  - \frac{ \sin{\theta}_{\nu} }{ \sqrt{2} }  & 
  \frac{ \cos{\theta}_{\nu} }{ \sqrt{2} } & 
  \frac{1}{ \sqrt{2} }
 \end{pmatrix}
 \quad \textrm{and} \quad 
 {\bf \Delta}_{\nu}^{0} = 
  \textrm{diag} \left(  m^{0}_{\nu_{1}}, m^{0}_{\nu_{2}}, m^{0}_{\nu_{3}} \right) .
\end{align}  
So, the matrix elements of ${\bf M}^{0}_{\nu}$ may be written in terms of neutrino mass 
eigenvalues and the $\theta_{\nu}$ angle as:
\begin{equation}\label{e6}
 \begin{array}{ll}
  m_{ee} = 
   \left( m^{0}_{\nu_{1}} \cos^{2} \theta_{\nu} 
   + m^{0}_{\nu_ {2}} \sin^{2} \theta_{\nu} \right)  , & 
  m_{e\mu} = 
   \frac{ 
    \sin{ 2\theta_{\nu} }  \left( m^{0}_{\nu_{2}} - m^{0}_{\nu_{1}} \right) 
   }{ 
    \sqrt{8} 
   }, \\
  m_{\mu\mu} + m_{\mu\tau} = 
  m^{0}_{\nu_{1}} \sin^{2}{\theta}_{\nu} + m^{0}_{\nu_{2}} \cos^{2}{\theta}_{\nu}, &
  m_{\mu\mu} - m_{\mu\tau} = m^{0}_{\nu_{3}}.
 \end{array} 
\end{equation}

Strictly speaking, in the present model, ${\bf M}_{\nu}$ does not possess the $\lra$ 
symmetry since $m_{e\mu} \neq m_{e\tau}$. This fact, actually, is crucial to 
get $\theta_{13} \neq 0^{\circ}$ and $\theta_{23} \neq 45^{\circ}$ in the PMNS 
matrix as we will see next.
Now, the ${\bf M}_{\nu}$ mass matrix will be diagonalized in a perturbative way, as 
follows: applying ${\bf U}^{0}_{\nu}$ to ${\bf M}_{\nu}$, we have
\begin{equation}\label{Eq:MMd}
 {\bf U}^{0 \dagger}_{\nu} {\bf M}_{\nu} {\bf U}^{0 \ast}_{\nu} = 
 {\bf \Delta}_{\nu}^{0}  + 
 \begin{pmatrix}
  0 & 0 & \frac{ \cos{\theta_{\nu}} }{ \sqrt{2} } \left( m_{e\tau} - m_{e\mu} \right) \\ 
  0 & 0 & \frac{ \sin{\theta_{\nu}} }{ \sqrt{2} } \left( m_{e\tau} - m_{e\mu} \right) \\ 
  \frac{ \cos{\theta_{\nu}} }{ \sqrt{2} } \left( m_{e \tau} - m_{e \mu} \right) &   
  \frac{ \sin{\theta_{\nu}} }{ \sqrt{2} } \left( m_{e \tau} - m_{e \mu} \right) & 0
 \end{pmatrix} . 
\end{equation}
The right side of the above expression contains the difference 
$m_{e \tau} - m_{e \mu}$ that breaks the $\mu-\tau$ symmetry. Then, this mass matrix will 
be considered as a perturbation to the ${\bf M}^{0}_{\nu}$ matrix. 
As a consequence, the $m^{0}_{\nu_{i}}$ physical neutrino masses will get a correction 
as well as the ${\bf U}^{0}_{\nu}$ neutrino mixing. 
Here, we define the dimensionless perturbation parameter as 
\begin{equation}
 \epsilon \equiv
 \frac{ \left( m_{e\tau}-m_{e\mu} \right) }{ m_{e\mu} },
\end{equation}
where $\vert \epsilon \vert\ll 1$ as hypothesis. To be more precise, we will consider a soft breaking of the $\mu-\tau$ symmetry such that $\vert \epsilon \vert\ll 0.3$. Then, quadratic contributions on $\epsilon$ will be neglected along the analysis.
In this way, the second matrix of right side of Eq.~(\ref{Eq:MMd}) is given as
\begin{align}\label{e4}
 {\bf M}^{\epsilon}_{\nu} = 
 \begin{pmatrix}
  0 & 0 & \frac{ \cos{ \theta_{\nu} } }{ \sqrt{2} } m_{e \mu}~\epsilon \\ 
  0 & 0 & \frac{ \sin{ \theta_{\nu} } }{ \sqrt{2} } m_{e \mu}~\epsilon \\ 
  \frac{ \cos{ \theta_{\nu} } }{ \sqrt{2} } m_{e\mu}~\epsilon & 
  \frac{ \sin{ \theta_{\nu} } }{ \sqrt{2} } m_{e\mu}~\epsilon & 0
\end{pmatrix}  \, .
\end{align}

In general, the ${\bf M}_{\nu}$ mass matrix may be diagonalized by means of a unitary 
transformation 
${\bf U}^{\dagger}_{\nu} {\bf M}_{\nu} {\bf U}^{\ast}_{\nu} = {\bf \Delta}_{\nu}$, where
${\bf \Delta}_{\nu} = \textrm{diag} \left( m_{ \nu_{1} }, m_{ \nu_{2} }, m_{ \nu_{3} } 
\right)$,
${\bf U}_{\nu} \approx {\bf U}^{0}_{\nu} {\bf U}^{\epsilon}_{\nu}$ in this latter 
the ${\bf U}^{0}_{\nu}$ matrix diagonalizes to ${\bf M}^{0}_{\nu}$, while 
${\bf U}^{\epsilon}_{\nu}$ makes the same for the resultant matrix that depends on 
$\epsilon$.
%
%
%
%
Here, the $m^{0}_{\nu_{i}}$ active neutrino masses are complex due to the 
presence of Majorana phases, and the $\theta_{\nu}$ angle is a free parameter. 
%
%
The explicit form of the unitary matrix ${\bf U}^{\epsilon}_{\nu}$ is the following
\begin{align}\label{e10}
{\bf U}^{\epsilon}_{\nu} \approx
 \begin{pmatrix}
  N_{1}^{-1} & 0 & \frac{ k_{1}r_{1} \epsilon }{ N_{3} } \\ 
  0 & N_{2}^{-1} & \frac{ k_{2}r_{2} \epsilon }{ N_{3} } \\ 
  - \frac{ k_{1} r_{1} \epsilon }{ N_{1} } & 
  - \frac{ k_{2} r_{2} \epsilon }{ N_{2} } & N_{3}^{-1}
 \end{pmatrix},
\end{align}
where 
\begin{equation}\label{ec:r12-k12}
 r_{ 1,2 } \equiv 
  \left( \frac{ 
   m^{0}_{\nu_{2}} - m^{0}_{\nu_{1}} 
  }{ 
   m^{0}_{\nu_{3}} - m^{0}_{\nu_{1,2}} 
  }\right), \quad 
 k_{1} \equiv 
  \frac{ \sin{ 2\theta_{\nu} } \cos{ \theta_{\nu} } }{ 4 } 
 \quad \textrm{and} \quad
 k_{2} \equiv 
  \frac{ \sin{ 2\theta_{\nu} } \sin{ \theta_{\nu} } }{ 4 }.
\end{equation}
The normalization factors in the ${\bf U}^{\epsilon}_{\nu}$ are
\begin{equation}
 \begin{array}{l}\vspace{2mm}
  N_{1} = 
   \sqrt{ 1 + \left| \epsilon k_{1} r_{1} \right|^{2} } , \quad
  N_{2} =
   \sqrt{ 1 + \left| \epsilon k_{2} r_{2} \right|^{2} } , \quad
  N_{3} =
   \sqrt{ 1 + \left| \epsilon \right|^{2}
   \left( \left|k_{1} r_{1} \right|^{2} + \left|k_{2} r_{2} \right|^{2}\right) } .
 \end{array}
\end{equation}
Finally, the lepton flavor mixing matrix is given as
${\bf V}\approx {\bf U}^{\dagger}_{\ell} {\bf U}^{0}_{\nu} {\bf U}^{\epsilon}_{\nu}$, 
where ${\bf U}_{\ell}$ for this theoretical framework is equal to the unity matrix, while 
${\bf U}^{0}_{\nu}$ and ${\bf U}^{\delta}_{\nu}$ matrices are given in 
Eqs.~(\ref{e5e}) and~(\ref{e10}), respectively. Thus, we obtain
%
%
%
\begin{equation}\label{PMNS}
 {\bf V} \approx 
  \left( \begin{array}{ccc} \vspace{2mm}
   \frac{ \cos \theta_{\nu} }{ N_{1} } & 
   \frac{ \sin \theta_{\nu} }{ N_{2} } &
   \frac{ \epsilon \cos^{2} \theta_{\nu} \sin 2 \theta_{\nu} r_{1} 
    \left( 1 + \frac{ r_{2} }{ r_{1} } \tan^{2} \theta_{\nu} \right) }{ 4 N_{3} }   
   \\ \vspace{2mm}
   - \frac{ 
    \sin \theta_{\nu} \left( 1 - \epsilon ( \cos^{2}{ \theta_{\nu} }/2) 
     r_{1}  \right) }{ \sqrt{2} N_{1} } & 
  \frac{ 
    \cos \theta_{\nu} \left( 1 + \epsilon  (\sin^{2} \theta_{\nu}/2) r_{2} \right)
   }{ 
    \sqrt{2} N_{2} 
   } & 
  - \frac{ 
    1 - \epsilon (\sin^{2} 2 \theta_{\nu}/8) r_{1}r_{2}}{ 
     \sqrt{2} N_{3} 
   } \\ \vspace{2mm}
  - \frac{ 
    \sin \theta_{\nu} \left( 1 + \epsilon (\cos^{2}{\theta_{\nu}}/2) r_{1}  \right) 
   }{ 
    \sqrt{2} N_{1} 
   }  & 
  \frac{ 
    \cos \theta_{\nu} \left( 1 - \epsilon  (\sin^{2} \theta_{\nu}/2) r_{2} \right)
   }{ 
    \sqrt{2} N_{2} 
   } & 
  \frac{ 
    1 + \epsilon (\sin^{2} 2 \theta_{\nu}/8) r_{1}r_{2}
   }{ 
     \sqrt{2} N_3 
   } 
 \end{array} \right) \; .
\end{equation}
In order to obtain the theoretical expressions for the flavor mixing angles,  we compare  
the magnitude of the entries in the mixing matrix $PMNS$  given in the above parametrization 
and the Standard parametrization~\cite{Barradas-Guevara:2017iyt}. Then, in this 
theoretical framework the reactor, solar and atmospheric mixing angles have the form: 
\begin{equation}\label{e14}
 \begin{array}{l}\vspace{2mm}   
  \sin^{2} \theta_{13} =  
   \left| {\bf V}_{13} \right|^{2} = 
   \dfrac{ 
    \left| \epsilon \right|^{2}  \sin^{2} 2 \theta_{\nu} \cos^{4} {\theta_{\nu}}  
    \left| r_{1} \right|^{2} 
    \left| 1 + \frac{ r_{2} }{ r_{1} } \tan^{2} \theta_{\nu}  \right|^{2} 
   }{
    16 N^{2}_{3}
   } ,
 \end{array}
\end{equation}
\begin{equation}\label{e12}
 \begin{array}{l}\vspace{2mm}
  \sin^{2} \theta_{23}  =  
   \dfrac{ 
    \left| {\bf V}_{23} \right|^{2} 
   }{
    1 - \left| {\bf V}_{13} \right|^{2} 
   } =
   \dfrac{ 1 }{ 2 N^{2}_{3}} 
   \dfrac{ 
     \left| 1 -\frac{\sin^{2}{2\theta_{\nu}}}{8} \epsilon r_{1} r_{2}\right|^{2} 
    }{ 
     1-\sin^{2}{\theta}_{13} 
    } ,
 \end{array}
\end{equation}
\begin{equation}\label{e13}
 \begin{array}{l}\vspace{2mm}
  \sin^{2} \theta_{12} = 
   \dfrac{ 
    \left| {\bf V}_{12} \right|^{2} 
   }{
    1 - \left| {\bf V}_{13} \right|^{2} 
   } = 
 \dfrac{ 1 }{N^{2}_{2}} 
  \dfrac{\sin^{2}{\theta_{\nu}} }{1-\sin^{2}{\theta}_{13}}  .
 \end{array}   
\end{equation} 
%
In the limit when $\epsilon$ goes to zero, one recovers the well known results of exact 
$\mu-\tau$ symmetry: $\theta_{13}=0^{\circ}$ and $\theta_{23}=45^{\circ}$. 
On the other hand, in order to figure out the mixing angles values that our model 
predicts, an analytic study will be done. 
To do this, we should keep in mind that $\vert \epsilon \vert \ll 1$, to be more 
precisely $\vert \epsilon \vert \leq 0.3$. With this in mind, the normalization factors should be of order 
of $1$. 
Then the expression for the solar mixing angle to leading order is
\begin{equation}\label{e11'}
  \sin^{2} \theta_{12} \approx \sin^{2} \theta_{\nu} 
  \quad \Rightarrow \quad \theta_{12} \approx \theta_{\nu}.
\end{equation}   
Thus, along the analytical study, we will consider that 
$\sin{\theta_{\nu}}\approx 1/\sqrt{3}$ which is a good approximation to the solar 
angle~\cite{Harrison2002167,Harrison2002163,Xing200285,Harrison200376}. 
The other two mixing angles take the form 
\begin{equation}\label{e12'}
 \begin{array}{l}
  \sin^{2} \theta_{23}  \approx  
   \dfrac{ 1 }{ 2 }
   \left| 1 - \dfrac{ \epsilon }{ 8 } \sin^{2}{ 2 \theta_{12} } r_{1} r_{2} \right|^{2}    
   \quad \textrm{and} \quad      
  \sin^{2} \theta_{13} \approx
   \dfrac{ \left| \epsilon \right|^{2} }{ 16 }  
    \sin^{2} 2 \theta_{12} \cos^{4} \theta_{12}  \left| r_{1} \right|^{2} 
    \left| 1 + \dfrac{ r_{2} }{ r_{1} } \tan^{2} \theta_{12}  \right|^{2} .
 \end{array}
\end{equation}
The explicit form of the $r_{1,2}$ parameters are given in Eq.~(\ref{ec:r12-k12}). 
Notice that the reactor angle as well as the atmospheric one depend strongly on the complex 
neutrino masses, so that the Majorana phases may be relevant to enhance or suppress those.

In order to show this last fact, the diagonal matrix, given in Eq.~(\ref{e5e}), can be written 
as ${\bf \Delta}^{0}_{\nu} = \textrm{diag} \left( \vert m^{0}_{\nu_{1}} \vert e^{i\alpha_{1}}, 
\vert m^{0}_{\nu_{2}} \vert e^{i\alpha_{2}}, \vert m^{0}_{\nu_{3}}\vert e^{i\alpha_{3}} 
\right)$ where $\alpha_{i}$ is the respective Majorana phase for each neutrino mass. In the 
following study, we will consider the case of CP parities ($0$ or $\pi$) in the Majorana 
phases and some combinations among them. Then, we end up having 
${\bf \Delta}^{0}_{\nu} = \textrm{diag} \left( \pm \vert m^{0}_{\nu_{1}} \vert, 
\pm \vert m^{0}_{\nu_{2}} \vert, \pm \vert m^{0}_{\nu_{3}} \vert \right)$. 
As we will see,
there are several cases where these CP parities play an important role to constrain the 
$\epsilon$ parameter and the lightest neutrino mass that accommodate the reactor and 
atmospheric angles. For the moment, none of the Majorana phases will be factorized so this analysis is 
equivalent up to one phase to the standard parametrization where two relative Majorana phases 
are taking into account.

In the last part of the work, as a particular prediction of this model, we will calculate the effective neutrino 
mass that comes from the neutrinoless double beta decay, the neutrino mass scale and the sum 
of the neutrino masses. These observables will be calculated in the framework CP parities and 
two relative Majorana phases; in the latter case, the analytical study is out of the scope of this work, but naive plots will be presented in order to compare them with the former case. 
For convenience, we will work with the parametrization 
${\bf \Delta}^{0}_{\nu} = 
\textrm{diag} \left( \vert m^{0}_{\nu_{1}} \vert e^{i\alpha}, \vert m^{0}_{\nu_{2}} \vert, 
\vert m^{0}_{\nu_{3}} \vert e^{i\beta} \right) e^{i\alpha_{2}}$ where 
$\alpha \equiv \alpha_{1}-\alpha_{2}$, $\beta\equiv\alpha_{3}-\alpha_{2}$ and the $\alpha_{2}$ 
phase is irrelevant.

Let us start with the analytical study in the framework of CP parities.
\begin{enumerate}
 \item {\bf Normal Hierarchy}. 
  From the definition of $\Delta m^{2}_{21}$ and $\Delta m^{2}_{31}$, one obtains the 
  absolute masses 
  $\vert m^{0}_{\nu_{3}} \vert = \sqrt{\Delta m^{2}_{31} + m^{0 2}_{\nu_{1}}}$ and 
  $\vert m^{0}_{\nu_{2}} \vert = \sqrt{\Delta m^{2}_{21} + m^{0 2}_{\nu_{1}}}$,
  \begin{align}
   r_{1} \approx 
    \frac{ m^{0}_{\nu_{2}} }{ m^{0}_{\nu_{3}} } 
    \left( 1 - \frac{ m^{0}_{\nu_{1}} }{ m^{0}_{\nu_{2}} } \right), 
    \qquad 
   \frac{r_{2}}{r_{1}} \approx 
    1 + \frac{ m^{0}_{\nu_{2}} }{ m_{\nu_{3}} }, 
   \qquad
   r_{1} r_{2} \approx 
    \frac{ m^{0}_{\nu_{2}} }{ m^{0}_{\nu_{3}} } 
    \left( 1 + \frac{ m^{0}_{\nu_{2}} }{ m^{0}_{\nu_{3}} } \right)
    \left( 1 - \frac{ m^{0}_{\nu_{1}} }{ m^{0}_{\nu_{2}} } \right)^{2}. \label{nh1}
  \end{align}
  The above mass ratios will depend on the sign of $m^{0}_{\nu_{2}}$ and  
  $m^{0}_{\nu_{3}}$. 
  At the same time $m^{0}_{\nu_{2}}/m^{0}_{\nu_{3}}\approx \mathcal{O}( \sqrt{ 
  \Delta m^{2}_{12}/\Delta m^{2}_{31}})$ up to some signs. 
  Therefore, the reactor angle will be proportional to 
  $(\Delta m^{2}_{12}/\Delta m^{2}_{31}) \vert \epsilon \vert^{2}$ which turns out being 
  tiny. 
  This statement holds to whatever signs are assumed in the neutrino masses, 
  $m^{0}_{\nu_{2}}$ and  $m^{0}_{\nu_{3}}$. Because of this, we may infer the normal case 
  is ruled out for $\vert \epsilon \vert \leq 0.3$.

\item{\bf Inverted Hierarchy}. 
Following a similar analysis to the normal case, we have $\vert m^{0}_{\nu_{2}} \vert 
= \sqrt{ \Delta m^{2}_{31} + \Delta m^{2}_{21} + \vert m^{0}_{\nu_{3}}\vert^{2}}$ and 
$\vert m^{0}_{\nu_{1}} \vert = \sqrt{\Delta m^{2}_{13} + \vert m^{0}_{\nu_{3}} 
\vert^{2} }$. For the mass ratios, we have

\begin{align}
r_{1}&\approx -\left( \frac{m^{0}_{\nu_{2}}-m^{0}_{\nu_{1}}}{m^{0}_{\nu_{1}}}\right)
\left(1+\frac{m^{0}_{\nu_{3}}}{m^{0}_{\nu_{1}}}\right),\quad \frac{r_{2}}{r_{1}}\approx 
\frac{m^{0}_{\nu_{1}}}{m^{0}_{\nu_{2}}}\left[1-\frac{m^{0}_{\nu_{3}}(m^{0}_{\nu_{2}}-
m^{0}_{\nu_{1}})}{m^{0}_{\nu_{2}}m^{0}_{\nu_{1}}}\right],\nonumber\\
r_{1}r_{2}&\approx \frac{(m^{0}_{\nu_{2}}-m^{0}_{\nu_{1}})^{2}}{m^{0}_{\nu_{1}}
m^{0}_{\nu_{2}}}
\left[1+\frac{m^{0}_{\nu_{3}}(m^{0}_{\nu_{2}}+m^{0}_{\nu_{1}})}{m^{0}_{\nu_{2}}
m^{0}_{\nu_{1}}}\right].\label{ih1}
\end{align}
 Notice that these mass ratios are sensitive at the sign of the neutrino masses 
  $m^{0}_{\nu_{i}}$. Actually, there are four independent scenarios: {\bf Case A} with   
  $m^{0}_{\nu_{i}}>0$; {\bf Case B} with $m^{0}_{\nu_{(1,2)}}>0$ and $m^{0}_{\nu_{3}}<0$; 
  {\bf Case C} with $m^{0}_{\nu_{(2,3)}}>0$ and $m^{0}_{\nu_{1}}< 0$; {\bf Case D} with 
  $m^{0}_{\nu_{2}}>0$ and $m^{0}_{\nu_{(1,3)}}<0$. The former cases can be written as
  \begin{align}
   r_{1} & \approx 
    - \left( \frac{ \left| m^{0}_{\nu_{2}} \right| - \vert m^{0}_{\nu_{1}} \vert 
    }{ 
     \vert m^{0}_{\nu_{1}} \vert} \right) \left( 1\pm \frac{ \left| m^{0}_{\nu_{3}}
     \right| }{ \left| m^{0}_{\nu_{1}} \right|} \right), \qquad 
   \frac{ r_{2} }{ r_{1} }\approx 
	\frac{ \left|m^{0}_{\nu_{1}} \right| }{ \left| m^{0}_{\nu_{2}} \right| } 
	\left[ 1 \mp
	\frac{ \left| m^{0}_{\nu_{3}} \right| \left( \left| m^{0}_{\nu_{2}} \right| 
	 - \left| m^{0}_{\nu_{1}} \right| \right) 
	}{
	 \left| m^{0}_{\nu_{2}} \right| \left| m^{0}_{\nu_{1}} \right|} \right]; \nn \\
   r_{1} r_{2} & \approx 
    \frac{ 
     ( \left| m^{0}_{\nu_{2}} \right| - \left| m^{0}_{\nu_{1}} \right| )^{2} 
    }{
     \left| m^{0}_{\nu_{1}}\right| \left| m^{0}_{\nu_{2}}\right|
    }
	\left[1 \pm
	\frac{ 
	 \left| m^{0}_{\nu_{3}} \right|
	 \left( \left| m^{0}_{\nu_{2}} \right| + \left| m^{0}_{\nu_{1}} \right| \right) 
	}{
	 \left| m^{0}_{\nu_{2}} \right| \left| m^{0}_{\nu_{1}} \right|
	} \right], \label{ih2}
  \end{align}
  where the upper (lower) sign corresponds to the {\bf Case A} ({\bf Case B}). 
  For the {\bf Case C} ({\bf Case D}) the corresponding sign is the upper (lower), 
  and this is given by
  \begin{align}
   r_{1} & \approx 
    \left( \frac{ \left| m^{0}_{\nu_{2}} \right| + \vert m^{0}_{\nu_{1}} \vert 
    }{ \vert m^{0}_{\nu_{1}} \vert } \right) 
    \left( 1 \mp \frac{ \left| m^{0}_{\nu_{3}} \right| }{ 
    \left| m^{0}_{\nu_{1}} \right| } \right), \qquad 
   \frac{ r_{2} }{ r_{1} } \approx- 
	\frac{ \left|m^{0}_{\nu_{1}}\right| }{ \left| m^{0}_{\nu_{2}} \right| } 
	\left[ 1 \pm
 	 \frac{ \left|m^{0}_{\nu_{3}}\right| ( \left| m^{0}_{\nu_{2}} \right| 
	 + \left| m^{0}_{\nu_{1}} \right| ) }{ \left| m^{0}_{\nu_{2}} \right| 
	 \left| m^{0}_{\nu_{1}} \right| } \right]; \nn \\
   r_{1} r_{2} & \approx - \frac{ ( \left| m^{0}_{\nu_{2}} \right| 
    + \left| m^{0}_{\nu_{1}} \right| )^{2} }{ \left| m^{0}_{\nu_{1}}\right| 
    \left| m^{0}_{\nu_{2}} \right| } \left[ 1 \mp \frac{ \left| m^{0}_{\nu_{3}} \right|
	( \left| m^{0}_{\nu_{2}} \right| - \left| m^{0}_{\nu_{1}} \right| ) }{ 
	\left| m^{0}_{\nu_{2}} \right| \left| m^{0}_{\nu_{1}} \right| } \right]. 
   \label{ih3}
  \end{align}
  From the absolute neutrino mass expressions, we notice that 
  $\vert m^{0}_{\nu_{2}} \vert \approx \vert m^{0}_{\nu_{1}} \vert ( 1 + R_{1} )$, 
  then
  \begin{align}
   \vert m^{0}_{\nu_{2}} \vert - \vert m^{0}_{\nu_{1}} \vert \approx 
    \vert m^{0}_{\nu_{1}} \vert R_{1}, \qquad 
   \vert m^{0}_{\nu_{2}} \vert + \vert m^{0}_{\nu_{1}} \vert \approx 
    2 \vert m^{0}_{\nu_{1}} \vert \left( 1 + \frac{ R_{1} }{2} \right),\qquad 
   \vert m^{0}_{\nu_{2}} \vert \vert m^{0}_{\nu_{1}} \vert \approx 
    \vert m^{0}_{\nu_{1}} \vert^{2} ( 1 + R_{1} ) ,
   \label{ih5d}
  \end{align}
  where $R_{1} \equiv \Delta m^{2}_{21}/2 \vert m^{0}_{\nu_{1}} \vert^{2} \sim 10^{-2}$ 
  which is valid whereas the lightest neutrino mass will be tiny.

  Roughly speaking, for the {\bf Cases A} and {\bf B}, the mixing angles are similar and 
  these are given as
  \begin{align}
   \sin^{2}{ \theta_{13} } \approx 
    \frac{ \vert \epsilon \vert^{2} }{ 18 } R^{2}_{1} 
    \left( 1 - \frac{2}{3} R_{1} \right), \qquad 
   \sin^{2}{ \theta_{23} } \approx 
    \frac{1}{2} \left| 1 - \frac{ \epsilon }{ 9 } R^{2}_{1} \right|^{2}.
  \end{align}
  These relations allow us to discard the {\bf Case A} and {\bf B} since the reactor 
  angle is proportional to $\vert \epsilon \vert^{2} R^{2}_{1}$ that turns out too
  small for $\vert \epsilon \vert \leq 0.3$. 
  In the {\bf Case C} and {\bf D}, one obtains
  \begin{align}
   \sin^{2} {\theta}_{13} & \approx 
    \frac{2}{81} \vert \epsilon \vert^{2} \left( 1 + 3 R_{1} \mp 6 
    \frac{ \vert m^{0}_{\nu_{3}} \vert }{ \vert m^{0}_{\nu_{1}} \vert} \right), \quad
   \sin^{2}{\theta_{23}} \approx 
    \frac{1}{2} \left| 1 + \frac{4}{9} \epsilon \left( 1 \mp 
    \frac{ \vert m^{0}_{\nu_{3}} \vert }{ \vert m^{0}_{\nu_{1}} \vert } 
    R_{1} \right) \right|^{2}. \label{ih6}
  \end{align}
  From these formulas, in the strict inverted hierarchy, 
  $\vert m^{0}_{\nu_{3}} \vert = 0$ then $R_{1} \rightarrow \Delta m^{2}_{21}/2 
  \Delta m^{2}_{13}$, we obtain:

  (a) If $\vert \epsilon \vert = 0.3$,
  \begin{align}
   \sin^{2}{ \theta}_{13} & \approx 0.0023, \quad
   \sin^{2}  \theta_{23} = 
    \left\{ \begin{array}{l} \vspace{2mm}
	 0.64, \quad \textrm{with} \quad \alpha_{\epsilon} = 0\\ 
     0.37, \quad \textrm{with} \quad \alpha_{\epsilon} = \pi
  \end{array}\right.
  \end{align}

  (b) If $\vert \epsilon \vert = 0.1$,
  \begin{align}
   \sin^{2}{\theta}_{13} & \approx 0.00025, \quad
   \sin^{2} \theta_{23}= 
    \left\{ \begin{array}{l} \vspace{2mm}
     0.54,\quad \textrm{with}\quad \alpha_{\epsilon} = 0\\ 
     0.45,\quad \textrm{with}\quad \alpha_{\epsilon} = \pi
    \end{array}\right.
  \end{align}
  In here, we emphasize the importance of the 
  $\vert \epsilon \vert e^{i \alpha_{\epsilon}}$ associated phase of the perturbation parameter since this might lead to a deviation above or below of $45^{\circ}$ the 
  atmospheric angle. 
  Then, from above cases the atmospheric angle is accommodated in the allowed region with 
  $\alpha_{\epsilon} = 0$. 
  Therefore, this value for the phase will be chosen when the lightest neutrino mass is very small but different from zero
 .

  Going back to Eq.~(\ref{ih6}), we observe that the {\bf Case C} will be disfavored 
  since the reactor angle is reduced by the term 
  $6 \vert m^{0}_{\nu_{3}} \vert / \vert m^{0}_{\nu_{1}} \vert$; this statement can be 
  verified in a straightforward way, as it will be done also for the {\bf Case D}. Fixing  
  the reactor angle to its central value we can figure out the allowed values for the 
  perturbation parameter and the lightest neutrino mass. 
  Then, if $\vert m^{0}_{\nu_{3}} \vert = 0.001$, one requires that 
  $\vert \epsilon \vert \approx 0.89$ to get $\sin^{2}{\theta}_{13} = 0.0229$. 
  As result, we obtain $\sin^{2}{ \theta_{23} } \approx 0.97$ which is large in 
  comparison to the allowed values. 
  Now, if $\vert m^{0}_{\nu_{3}}\vert= 0.1$,  we need that 
  $\vert \epsilon \vert\approx 0.38$ to get $\sin^{2} {\theta}_{13} = 0.0229$, in this 
  way,  $\sin^{2} {\theta_{23}} \approx 0.68$ which is on the top of the allowed 
  experimental region.
  Remarkably, in this brief analysis the value of the perturbation parameter and the lightest neutrino mass are
  on the limit of soft breaking of the $\mu-\tau$ symmetry and the degenerate region, 
  respectively. The second plot of figure {\ref{fi2}} shows how $\vert m^{0}_{\nu_{3}} \vert$ depends on $\vert \epsilon \vert$ and viceversa, which is consistent with the previous analysis. In addition,  notice that the reactor angle prefers large values for 
  $\vert \epsilon \vert$ whereas the atmospheric one is favored with small values.

  In order to figure out the parameter space that fits the best values of the reactor and 
  atmospheric angles, the exact formulas for these were taken and the observables as 
  $\Delta m^{2}_{21}$, $\Delta m^{2}_{13}$ and $\theta_{\nu}$ were considered up to 
  $3\sigma$. With this in mind, in the following plots for the inverted and degenerate hierarchy, we demand that the reactor angle within $3~\sigma$ of C. L. of its experimental values (see the first plot in figure \ref{fi4}) to determine and constrain the atmospheric angle and the free parameters $\vert \epsilon \vert$ and $\vert m^{0}_{\nu_{3}}\vert$, respectively. The set of values for these parameters are shown in the next plots.
  
  \begin{figure}[ht]
   \centering
   \includegraphics[scale=0.6]{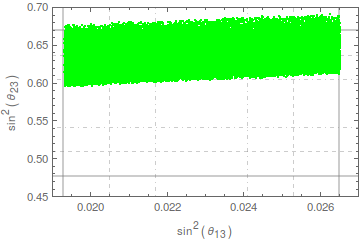}\hspace{0.5cm}\includegraphics[scale=0.6]{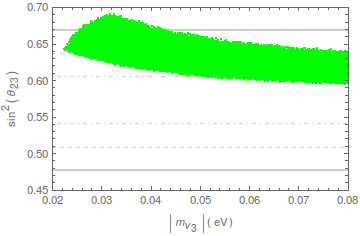}
   \caption{
    {\bf Case D}: Allowed region for $\sin^{2}{\theta_{13}}$ and $\sin^{2}{\theta_{23}}$, 
    respectively. The dotdashed, dashed and thick lines stand for $1~\sigma$, $2~\sigma$ and $3~\sigma$, respectively} \label{fi4}
  \end{figure}

\begin{figure}[ht]
\centering
\includegraphics[scale=0.6]{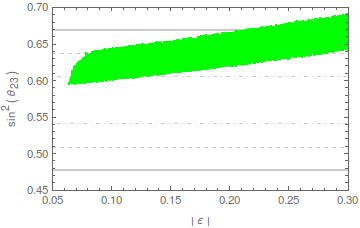}\hspace{0.5cm}\includegraphics[scale=0.58]{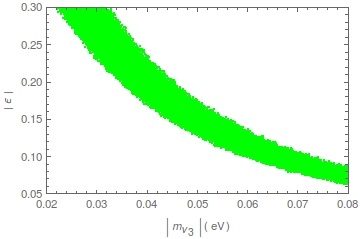}
\caption{
{\bf Case D}: Allowed region for $\sin^{2}{\theta_{23}}$ versus $\vert \epsilon \vert$ and $\vert \epsilon \vert$ versus $\vert m^{0}_{\nu_{3}} \vert$. The dotdashed, dashed and thick lines stand for $1~\sigma$, $2~\sigma$ and $3~\sigma$, respectively}\label{fi2}
  \end{figure}
 \item {\bf Degenerate Hierarchy}. 
  In this case, we have that 
  $\vert m^{0}_{\nu_{1}} \vert \approxeq \vert m^{0}_{\nu_{2}} \vert \approxeq 
  \vert  m^{0}_{\nu_{3}} \vert \approxeq m_{0}$ with $m_{0} \gtrsim 0.1~eV$. 
  Then, the absolute neutrino masses are 
  $\vert m^{0}_{\nu_{3}} \vert = \sqrt{ \Delta m^{2}_{31} + m^{2}_{0} } \approx 
  m_{0} \left( 1 + \Delta m^{2}_{31} / 2m^{2}_{0} \right)$ and 
  $\vert m^{0}_{\nu_{2}} \vert = \sqrt{ \Delta m^{2}_{21} + m^{2}_{0}} \approx 
  m_{0} \left( 1 +\Delta m^{2}_{21}/2m^{2}_{0} \right)$. 
  As in the inverted case, the extreme Majorana phases in the neutrino masses are 
  relevant for the mixing angles, as we will see next. There are four independent cases for 
  the signs which are shown below.
  \begin{itemize}
   \item {\bf Case A}. 
    If $m^{0}_{\nu_{i}} > 0$, then we have that
    \begin{align}
     r_{1} = 
      \frac{ 
       \vert m^{0}_{\nu_{2}} \vert - m_{0} 
      }{ 
       \vert m^{0}_{\nu_{3}} \vert - m_{0} 
      }, \quad 
     r_{1} r_{2} = 
      \frac{ 
       ( \vert m^{0}_{\nu_{2}} \vert -m_{0} )^{2} 
      }{ 
       ( \vert m^{0}_{\nu_{3}} \vert - \vert m^{0}_{\nu_{2}} \vert ) 
       ( \vert m^{0}_{\nu_{3}} \vert - m_{0} ) 
      }, \quad 
     \frac{ r_{2} }{ r_{1} } = 
      \frac{ 
       \vert m^{0}_{\nu_{3}} \vert - m_{0} 
       }{
        \vert m^{0}_{\nu_{3}} \vert - \vert m^{0}_{\nu_{2}} \vert}.\label{dh1}
    \end{align}
   \item {\bf Case B}. 
    If $m^{0}_{\nu_{(1,2)}} > 0$ and $m^{0}_{\nu_{3}} < 0$. As result of this, we get
    \begin{align}
	 r_{1} = 
	  -\frac{ 
	   \vert m^{0}_{\nu_{2}} \vert - m_{0} 
	  }{
	   \vert m^{0}_{\nu_{3}} \vert + m_{0} 
	  }, \quad 
	 r_{1} r_{2} = 
	  \frac{ 
	   ( \vert m^{0}_{\nu_{2}} \vert - m_{0} )^{2} 
	  }{ 
	   ( \vert m^{0}_{\nu_{3}} \vert + \vert m^{0}_{\nu_{2}} \vert) 
	   ( \vert m^{0}_{\nu_{3}} \vert + m_{0}) 
	  }, \quad 
	 \frac{ r_{2} }{ r_{1} } = 
	  \frac{ \vert m^{0}_{\nu_{3}} \vert + m_{0 }}{ \vert m^{0}_{\nu_{3}} \vert 
	  + \vert m^{0}_{\nu_{2}} \vert}.\label{dh4}
	\end{align}
   \item {\bf Case C}. 
    If $m^{0}_{\nu_{(2,3)}} > 0$ and $m^{0}_{\nu_{1}} = -m_{0}$, therefore one obtains
    \begin{align}
     r_{1} = 
      \frac{ 
       \vert m^{0}_{\nu_{2}} \vert + m_{0} 
      }{ 
       \vert m^{0}_{\nu_{3}} \vert + m_{0} 
      }, \quad 
     r_{1} r_{2} = 
      \frac{ 
       ( \vert m^{0}_{\nu_{2}} \vert + m_{0} )^{2} 
      }{ 
       ( \vert m^{0}_{\nu_{3}} \vert - \vert m^{0}_{\nu_{2}} \vert )
       ( \vert m^{0}_{\nu_{3}} \vert + m_{0} )
      }, \quad 
     \frac{ r_{2} }{ r_{1} } = 
      \frac{ 
       \vert m^{0}_{\nu_{3}} \vert + m_{0} 
      }{ 
       \vert m^{0}_{\nu_{3}} \vert - \vert m^{0}_{\nu_{2}} \vert 
      }. \label{dh2}
    \end{align}
   \item {\bf Case D}. 
    If $m^{0}_{\nu_{2}} > 0$ and $m^{0}_{\nu_{(1, 3)}} < 0$, then we have
    \begin{align}
     r_{1} = 
      - \frac{ 
       \vert m^{0}_{\nu_{2}} \vert + m_{0} 
      }{ 
       \vert m^{0}_{\nu_{3}} \vert- m_{0} 
      }, \quad 
     r_{1} r_{2} = 
      \frac{ 
       ( \vert m^{0}_{\nu_{2}} \vert + m_{0} )^{2} 
      }{ 
       ( \vert m^{0}_{\nu_{3}} \vert + \vert m^{0}_{\nu_{2}} \vert ) 
       ( \vert m^{0}_{\nu_{3}} \vert - m_{0} ) 
      }, \quad 
     \frac{r_{2}}{r_{1}} = 
      \frac{ 
       \vert m^{0}_{\nu_{3}} \vert - m_{0} 
      }{ 
       \vert m^{0}_{\nu_{3}} \vert + \vert m^{0}_{\nu_{2}} \vert 
      }.\label{dh3}
    \end{align}
  \end{itemize}
\end{enumerate}
Now, observe that
\begin{equation}\label{dh5}
 \begin{array}{ll}
  \vert m^{0}_{\nu_{2}} \vert - m_{0} \approx 2 m_{0} R_{2}, &
  \vert m^{0}_{\nu_{2}} \vert + m_{0} \approx 2 m_{0} \left( 1 + R_{2} \right), \\
  \vert m^{0}_{\nu_{3}} \vert - m_{0} \approx 2 m_{0} R_{3}, &
  \vert m^{0}_{\nu_{3}} \vert + m_{0} \approx 2 m_{0} \left( 1 + R_{3} \right), \\ 
  \vert m^{0}_{\nu_{3}} \vert - \vert m^{0}_{\nu_{2}} \vert \approx 
   2 m_{0} ( R_{3} - R_{2} ), &
  \vert m^{0}_{\nu_{3}} \vert + \vert m^{0}_{\nu_{2}} \vert \approx 
   2 m_{0} \left[ 1 + R_{2} + R_{3} \right]. 
 \end{array}
\end{equation}
Here, $R_{2} = \Delta m^{2}_{21}/4m^{2}_{0}$ and $R_{3} = \Delta m^{2}_{31}/4m^{2}_{0}$, 
then, $R_{2} \ll R_{3}$. 
From these expressions, in the {\bf Case A} and {\bf B}, we obtain respectively
\begin{align}
 \sin^{2}{ \theta_{13} } & 
  \approx \frac{ \vert \epsilon \vert^{2} }{ 18 } 
  \left( \frac{ R_{2} }{ R_{3} } \right)^{2} 
  \left( 1 + \frac{2}{3} \frac{ R_{2} }{ R_{3} } \right),\qquad 
 \sin^{2}{ \theta_{23} } \approx 
  \frac{1}{2} \left| 1 - \frac{ \epsilon }{ 9 } \frac{ R_{2} }{ R_{3} } \right|^{2} 
  \nonumber\\
 \sin^{2}{ \theta_{13} } & 
  \approx \frac{ \vert \epsilon \vert^{2} }{ 18 } R^{2}_{2} 
  \left[ 1 + \frac{2}{3} ( R_{2} + 2 R_{3} ) \right],\qquad 
 \sin^{2}{ \theta_{23} } \approx 
  \frac{1}{2} \left| 1 - \frac{\epsilon}{9} R^{2}_{2} \right|^{2}.
\end{align}
Both cases are discarded since the reactor angle is proportional to the small 
quantities $( R_{2} / R_{3} )^{2} \approx \mathcal{O} ( 10^{-3} )$ and 
$( R_{2} )^{2} \approx \mathcal{O}(10^{-5})$, respectively. 
Here, we have taken the central values for the inverted hierarchy and 
$m_{0} \approx 0.1~eV$, in addition $1/R_{3}\approx 16$ and 
$\vert \epsilon \vert \leq 0.3$. 
In fact, the values of $R_{2}$ and $R_{3}$ might be tiny if $m_{0}$ is large.

In the {\bf  Cases C} and {\bf D}, we have respectively
\begin{align}
 \sin^{2}{ \theta_{13} } & 
  \approx \frac{ \vert \epsilon \vert^{2} }{ 18 } 
  \left( 1 + 2R_{2} - 2R_{3} \right) 
  \left( 1 + \frac{1}{3 R_{3} } \right)^{2}, \quad
 \sin^{2}{ \theta_{23} }\approx 
  \frac{1}{2} \left| 1 - \dfrac{ \epsilon }{ 9 } 
  \left( \frac{ 1 + R_{2} - R_{3} }{ R_{3} } \right) \right|^{2}, \nonumber\\
 \sin^{2}{ \theta_{13} } & 
  \approx \frac{2}{81} \vert \epsilon \vert^{2}  
  \left( 1 + R_{3} \right) \left( \frac{ 1 + R_{2} }{ R_{3} } \right)^{2}, \quad
 \sin^{2}{ \theta_{23} } \approx 
  \frac{1}{2} \left| 1 - \dfrac{ \epsilon }{ 9 } 
  \left( \frac{ 1 + R_{2} - R_{3} }{ R_{3} } \right) \right|^{2}.
 \label{dh7}
\end{align}
As can be noticed, in the {\bf Case D}, the reactor angle is proportional to 
$(1/R_{3})^{2}$, which is a large quantity so that one requires that $\vert \epsilon \vert $ should 
be $10^{-2}$ to obtain the region where the atmospheric angle lies on. 
This favors the atmospheric angle since small values of $\vert \epsilon \vert$ are 
needed to not deviate too much from $45^{\circ}$. On the contrary, in the {\bf Case C} 
the atmospheric angle is deviated considerably since the $\vert \epsilon \vert$ value is 
large, in comparison to the above case, thus it is not necessary to suppress too 
much the $(1+1/3R_{3})^{2}$ factor in the reactor angle.

In a similar way to the inverted hierarchy, fixing the reactor angle to its central value 
we obtain the following for the {\bf Case C}:

 (a) If $m_{0}=0.1~eV$
 \begin{align}
  \sin^{2}{ \theta }_{13} & = 0.0229,\quad
  \sin^{2} \theta_{23} \approx
   \left\{ \begin{array}{l}\vspace{2mm}
    0.33, \quad \textrm{with} \quad \alpha_{\epsilon} = 0, \quad \textrm{and} \quad 
     \vert \epsilon \vert \approx 0.1 \\ 
    0.698, \quad \textrm{with} \quad \alpha_{\epsilon} = \pi, \quad \textrm{and} \quad 
     \vert \epsilon \vert \approx 0.1 
   \end{array} \right.
 \end{align}

 (b) If $m_{0}=0.25~eV$
 \begin{align}
  \sin^{2}{\theta}_{13} & = 0.0229,\quad
  \sin^{2} \theta_{23} \approx
   \left\{ \begin{array}{l} \vspace{2mm}
    0.31, \quad \textrm{with} \quad \alpha_{\epsilon} = 0, \quad \textrm{and} \quad 
     \vert \epsilon \vert \approx 0.018 \\ 
    0.73, \quad \textrm{with} \quad \alpha_{\epsilon} = \pi, \quad \textrm{and} \quad 
     \vert \epsilon \vert \approx 0.018
   \end{array}\right.
 \end{align}

In the {\bf Case D}, we obtain

 (a) If $m_{0}=0.1~eV$
 \begin{align}
  \sin^{2}{\theta}_{13} & = 0.0229, \quad
  \sin^{2} \theta_{23} \approx
   \left\{ \begin{array}{l}\vspace{2mm}
    0.41, \quad \textrm{with} \quad \alpha_{\epsilon} = 0, \quad \textrm{and} \quad 
     \vert \epsilon \vert \approx 0.055\\ 
    0.60, \quad \textrm{with} \quad \alpha_{\epsilon} = \pi,\quad \textrm{and} \quad 
     \vert \epsilon \vert \approx 0.055
   \end{array}\right.
 \end{align}

 (b) If $m_{0}=0.25~eV$
 \begin{align}
  \sin^{2}{\theta}_{13} & = 0.0229, \quad
  \sin^{2} \theta_{23} \approx
   \left\{ \begin{array}{l} \vspace{2mm}
    0.40, \quad \textrm{with} \quad \alpha_{\epsilon} = 0, \quad \textrm{and} \quad 
     \vert \epsilon \vert \approx 0.009 \\ 
    0.61, \quad \textrm{with} \quad \alpha_{\epsilon} = \pi, \quad \textrm{and} \quad 
     \vert \epsilon \vert \approx 0.009
   \end{array} \right.
\end{align}

\begin{figure}[ht]
 \centering
 \includegraphics[scale=0.6]{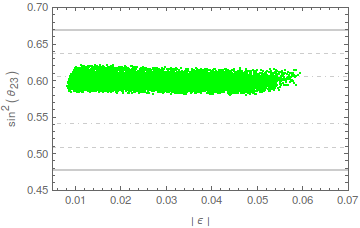}\hspace{0.5cm}\includegraphics[scale=0.57]{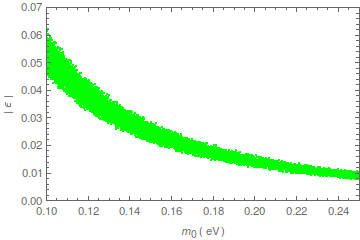}
 \caption{ 
  {\bf Case D}: Allowed region for $\sin^{2}{\theta_{23}}$ versus $\vert \epsilon \vert$ and $\vert \epsilon \vert$ versus $m_{0}$. The dotdashed, dashed and thick lines stand for $1~\sigma$, $2~\sigma$ and $3~\sigma$, respectively}\label{fi1}
\end{figure}

From the above results, we point out the importance of the $\alpha_{\epsilon}$ phase of 
the parameter $\epsilon$; this has to be $\pi$ in order to reach the allowed region for 
the atmospheric angle. Analogously to the inverted ordering, on the other hand, those results show the dependence between $m_{0}$ and $\vert \epsilon \vert$, the full allowed region is shown in the second plot of figure \ref{fi1}.

To get a panoramic view of the parameter space, the exact formulas for the reactor and atmospheric angles were taken for the {\bf Case D} and the observables as $\Delta m^{2}_{21}$,  $\Delta m^{2}_{13}$ and $\theta_{\nu}=\theta_{12}$ were considered up to $3~\sigma$. For the free parameters $\vert \epsilon \vert$ and $m_{0}$, the set of values is shown in the following plots. As we can see, the reactor and atmospheric angles are accommodated in good agreement with their experimental values.

\begin{figure}[ht]
 \centering
 \includegraphics[scale=0.6]{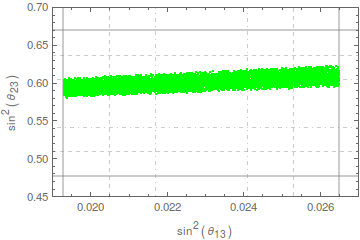}\hspace{0.5cm}\includegraphics[scale=0.6]{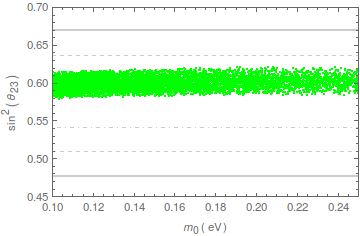}
 \caption{ 
  {\bf Case D}: Allowed region for $\sin^{2}{\theta_{13}}$ and $\sin^{2}{\theta_{23}}$, 
  respectively. The dotdashed, dashed and thick lines stand for $1~\sigma$, $2~\sigma$ and $3~\sigma$, respectively for each case.}\label{fi3}
\end{figure}

%
\section{Prediction on the absolute neutrino mass scale}
%
From the neutrino oscillation experiments, we get information on the mass square 
differences, but these experiments cannot say anything about the absolute neutrino mass scale. 
However, there are three processes that can address directly the determination of this 
important parameter: 
$i)$~analysis of CMB temperature fluctuations~\cite{Lesgourgues2006307},
$ii)$~the single $\beta$ decay~\cite{Otten086201} and 
$iii)$~neutrinoless double beta decay ($0\nu\beta\beta$)~\cite{RevModPhys.80.481}. 
The first one is purely observational and the neutrino masses are 
determined with kinematical methods.  The quantity probed in this approach is the sum of 
active neutrino masses, which we can denote as:
\begin{equation}
 m_{\mathrm{cosm}} = \sum_{i=1}^{3} m_{\nu_{i}} .
\end{equation}
An upper bound of $m_{\mathrm{cosm}}$ has been estimated by the {\bf Planck} collaboration, and 
its value is 
$m_{\mathrm{cosm}} = \sum_{i} \vert m_{\nu_{i}} \vert=0.23~eV$~\cite{Ade:2015xua}.
However, this method is called {\it model dependent} because it has a strong dependence on 
the cosmological and astrophysical assumptions~\cite{Lesgourgues2006307}.

On the other hand, the second and third processes for the determination of absolute neutrino mass scale, which we will study in this work, are based on 
the searches conducted in ground laboratories~\cite{Otten086201,RevModPhys.80.481}. 
In the $\beta$ decay processes, for example in the tritium decay, the effective electron mass 
is defined as~\cite{Giunti:Book,Osipowicz:2001sq,PhysRevD.76.113005,PhysRevD.73.053005, Rodejohann:2012xd,BILENKY-GIUNTI}: 
\begin{equation}\label{mef1}
 m_{\beta}^{2} = \sum_{i}^{3} m_{\nu_{i}}^{2} \left| V_{ei} \right|^{2},
\end{equation}
where $V_{ei}$ are the entries of lepton mixing matrix and correspond to the its first row.  
For a quasidegenerate neutrino mass spectrum 
$m_{\nu_{1}} \simeq m_{\nu_{2}} \simeq m_{\nu_{3}} \simeq m_{\nu_{e}}$, 
and for both possible hierarchies in the spectrum, the $m_{\beta}$ coincides approximately 
with the values of the three neutrino masses. 
Thus, Eq.~(\ref{mef1}) takes the form: 
$m_{\beta}^{2} \simeq m_{\nu_{e}}^{2} \sum_{i}^{3} \left| V_{ei} \right|^{2} = 
 m_{\nu_{e}}^{2}$. 
The Mainz~\cite{BONN2002133}, Troitsk~\cite{PhysRevD.84.112003} and
KATRIN~\cite{Katrin-Rep-2004} experiments can give us information on the absolute values of 
neutrino masses in the quasidegenerate region. From the first two experiments we have 
obtained  an upper limit to the electron anti-neutrino mass of 2.3 eV. The KATRIN 
experiment was designed to improve this limit by one order of magnitude down, such that either it discovers its mass, or sets an upper limit of $0.2$eV~\cite{Katrin-Rep-2004,Otten:2008zz,Capozzi:2017ipn}.

A direct measurement of the absolute neutrino mass scale through $\beta$ decay experiments, will 
not give us any information on the Dirac or Majorana character of neutrinos. 
The above issue can be solved by means of the neutrinoless  double beta decay experiments
($0\nu\beta\beta$). This is because the $0\nu\beta\beta$ decay is only possible if neutrinos 
are Majorana particles.
Also, with this decay process we can probe the absolute neutrino  mass scale by measuring of 
the effective Majorana mass of the electron neutrino, which is defined as:
\begin{equation}\label{mef}
 \vert m_{ee} \vert =
 \left| \sum_{i=1}^{3} m_{\nu_{i}} V_{ei}^{2}  \right|.
\end{equation} 
The lowest upper bound on $\vert m_{ee}\vert$ is provided by {\bf GERDA} phase-I 
data~\cite{Agostini:2013mzu}, and this is $0.22~eV$. 
That value will be substantially reduced by {\bf GERDA} phase-II 
data~\cite{Agostini:2017iyd}.


Now, for our theoretical framework and in the case of CP parities, 
the {\bf Case D} is the most favorable to accommodate the reactor and atmospheric angles for 
the inverted and degenerate hierarchy. Therefore, the above observables will be studied in 
this case. We have to keep in mind that the parameter that breaks the $\mu-\tau$ 
symmetry is small in both hierarchies so that the normalization factors are 
$N_{i}\approx \mathcal{O}(1)$. Due to CP parities the $\vert m_{ee}\vert$ effective neutrino 
mass becomes a real quantity, additionally there will be a cancellation among the involved 
terms; as a consequence, we expect small values in comparison to the {\bf GERDA} phase-I 
data.

In the scatter plots that will be shown later, for certain quantities the green and red colors stand 
for the CP parities and the CP non trivial values for the Majorana phases, respectively. In order to 
get these plots, the previous method and the same conditions in determining the atmospheric angle and constraining the $\vert \epsilon\vert $ parameter and the lightest neutrino mass were used. This is, the exact formulas for the respective quantities are taken into account, 
additionally the observables as $\Delta m^{2}_{21}$, $\Delta m^{2}_{13}$ and $\theta_{12}$ 
were considered up to $3\sigma$ of C.L. At the same time, the reactor angle was fixed to be consistent with the experimental value up to $3~\sigma$ of C. L. In this manner, we have calculated naively the following quantities in the presence of CP parities and CP non trivial values for the Majorana phases.

\begin{enumerate}
 \item{\bf Inverted Hierarchy}. The effective neutrino mass $\vert m_{ee}\vert$ is given by
  \begin{align}
   \left|m_{ee} \right|  \approx 
   \frac{1}{3} \vert m^{0}_{\nu_{1}}\vert \left|1+3 \left| 
   \frac{m^{0}_{\nu_{3}}}{ m^{0}_{\nu_{1}}} \right| \sin^{2}{\theta_{13}}- R_{1} \right|.
  \label{meih}
  \end{align}
  where the Eq.~(\ref{PMNS}) has been used, 
  $R_{1} = \Delta m^{2}_{21} / 2 \vert m^{0}_{\nu_{1}} \vert^{2}$ and 
  $\sin{\theta_{\nu}} \approx 1/\sqrt{3}$. In the limit case of strict inverted ordering, 
  $\vert m^{0}_{\nu_{3}}\vert=0$, one would have a defined value so we will expect that
  \begin{align}
   \left| m_{ee} \right| > 
   \frac{ \sqrt{ \Delta m^{2}_{13}} }{ 3 } 
   \left[ 1 - \frac{1}{2} \frac{ \Delta m^{2}_{21} }{ \Delta m^{2}_{13} } 
   \right] \approx 0.016~eV.
  \end{align}
  Since the lightest neutrino mass is not allowed to be zero as can be seen in 
  Fig.~\ref{fi4}. According of the allowed values for the lightest neutrino mass and the 
  $\epsilon$ parameter, the complete region of values for $\vert m_{ee}\vert$ is displayed in 
  Fig.~\ref{mee1}. In these plots, we have considered that $\alpha_{\epsilon}=0$ as was shown 
  in the analytic study.
  \begin{figure}[ht]\centering
   \includegraphics[scale=0.6]{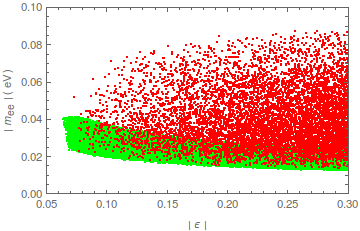}\hspace{0.5cm}\includegraphics[scale=0.6]{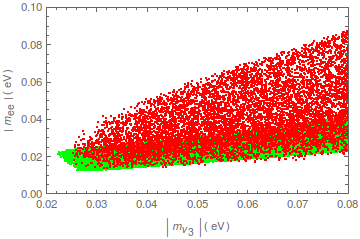}
   \caption{Allowed region for $\vert m_{ee}\vert$ versus $\vert m^{0}_{\nu_{3}}\vert$ and 
   $\vert \epsilon \vert$, respectively.}\label{mee1}
  \end{figure}
  In the case of CP parities, the predicted region of values are below the lowest upper 
  bound~\cite{Agostini:2013mzu}. 
  However, in the presence of CP non trivial values for the relative Majorana phases, 
  $\alpha$ and $\beta$, the allowed region may be increased up to the region where 
  {\bf GERDA} phase-II data will be sensitive.

  The neutrino mass scale in $beta$ decays and the sum of neutrino masses are given 
  respectively as
  \begin{align}
   m_{\nu_{e}} \approx \vert m^{0}_{\nu_{1}} \vert \left[ 1 + \frac{1}{3} R_{1} 
   + \frac{1}{2} \left| \frac{ m^{0}_{\nu_{3}} }{ m^{0}_{\nu_{1}} } \right|^{2} 
   \sin^{2}{ \theta_{13} } \right], \qquad 
   \sum_{i} \vert m_{\nu_{i}} \vert \approx 
   2 \vert m^{0}_{\nu_{1}} \vert \left[ 1 + \frac{1}{2} \left( R_{1} + \left| 
   \frac{ m^{0}_{\nu_{3}} }{ m^{0}_{\nu_{1}}} \right|  \right) \right] .
  \end{align}
  These observables are expected to be large than the limit values, 
  $\vert m^{0}_{\nu_{3}} \vert=0$, this means
  \begin{align}
   m_{\nu_{e}} > \sqrt{ \Delta m^{2}_{13} } \left[ 1 + \frac{1}{6} 
   \frac{ \Delta m^{2}_{21} }{ \Delta m^{2}_{13} } \right] \approx 0.049~eV, \qquad 
   \sum_{i} \vert m_{\nu_{i}} \vert > 2 \sqrt{ \Delta m^{2}_{13} } 
   \left[ 1 + \frac{1}{2} \frac{ \Delta m^{2}_{21} }{ \Delta m^{2}_{13} } \right] 
   \approx 0.098~eV.
  \end{align}
  The central values for the $\Delta m^{2}_{21}$, $\Delta m^{2}_{31}$ and the fixed reactor 
  angle have been considered for this purpose. 
  Notice the neutrino mass scale is below the value reported, $m_{\nu_{e}} = 0.2~eV$, 
  as can be seen in Fig.~\ref{mee2}, in the case of CP parities and CP non trivial values. 
  However, the bound on the sum of the neutrino masses can be reached in both frameworks.
  \begin{figure}[ht]\centering
   \includegraphics[scale=0.6]{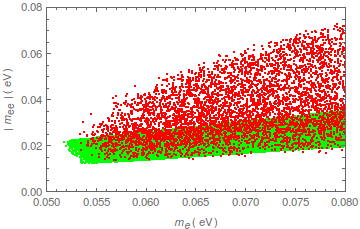}\hspace{0.5cm}\includegraphics[scale=0.6]{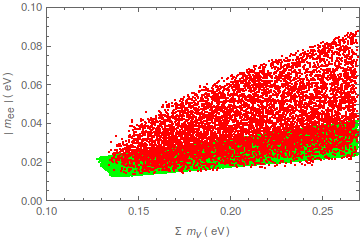}
   \caption{Predicted region for $\vert m_{ee} \vert$ versus the neutrino mass scale 
    $m_{\nu_{e}}$ and the sum of neutrino masses. }\label{mee2}
  \end{figure}
  
  To close this brief analysis, let us add two plots. The first one shows the allowed regions   
  for the atmospheric angle against the relative Majorana phases $\alpha$ and $\beta$. 
  According the analytical study, in the case of CP parities, the {\bf Case D} is the most 
  favorable to accommodate the mixing angles; this corresponds to $\alpha, \beta=\pi$. Along 
  with this, in the second panel the effective neutrino mass is shown against the two 
  relative Majorana phases. 
  These plots are consistent with our analytical results obtained previously. 
  \begin{figure}[ht]\centering
   \includegraphics[scale=0.6]{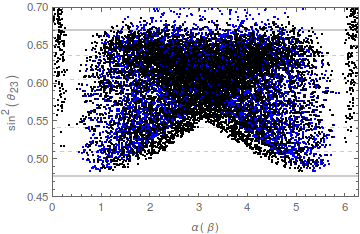}\hspace{0.5cm}\includegraphics[scale=0.6]{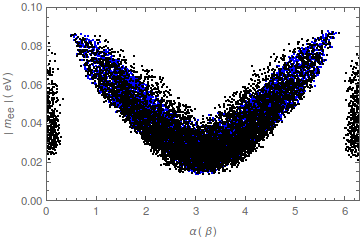}
   \caption{Blue and black points stand for $\alpha$ and $\beta$ phases, respectively.}
   \label{mee5}
  \end{figure}
 \item {\bf Degenerate Hierarchy}. In this case, we obtain
  \begin{align}
   \left| m_{ee} \right|  \approx 
   \frac{ m_{0} }{ 3 } \left| 1 + 3 \sin^{2}{ \theta_{13} }( 1 + 2 R_{3} ) - 2 R_{2} \right|.
  \end{align}
  Here the Eq.~(\ref{PMNS}) was used, $R_{2} = \Delta m^{2}_{21} / 4m^{2}_{0}$ and 
  $R_{3} = \Delta m^{2}_{13} / 4m^{2}_{0}$. With $m_{0} \approx 0.1~eV$, the lowest value for 
  the effective neutrino mass may be obtained, then $\vert m_{ee} \vert\geq 0.036~eV$ for 
  $m_{0}\gtrsim 0.1~eV$. The predicted regions for $\vert m_{ee} \vert$ can be seen in 
  Fig.~\ref{mee3}, in here, let us point out that the parameter space for $m_{0}$ and 
  $\vert \epsilon \vert$ is displayed in Figs.~\ref{fi3} and~\ref{fi1} where the 
  $\alpha_{\epsilon} = \pi$ phase was favored according to the analytical study.
  \begin{figure}[ht]\centering
   \includegraphics[scale=0.6]{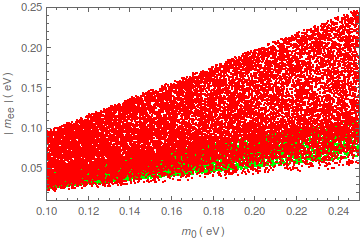}\hspace{0.5cm}\includegraphics[scale=0.6]{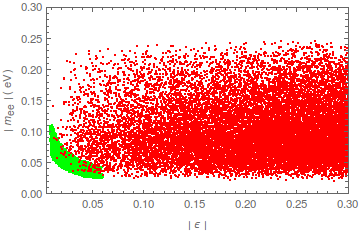}
   \caption{Allowed region for $\vert m_{ee}\vert$ versus $\vert m^{0}_{\nu_{3}}\vert$ and 
    $\vert\epsilon \vert$, respectively. }\label{mee3}
  \end{figure}

  Let us stress the following, in the case of CP parities, the breaking of the 
  $\mu-\tau$ symmetry was notable in that very small values for the $\vert \epsilon \vert$ parameter 
  were predicted by the model. Including two relative Majorana phases this soft breaking is 
  spoiled as can be seen in Fig.~(\ref{mee3}). 
  In a similar way to the inverted spectrum, the model predicts small values for 
  $\vert m_{ee}\vert$ in comparison to the {\bf GERDA} phase-I data, but the allowed region 
  is enhanced by the presence of CP non trivial values in the Majorana phases.

  Taking into account the $m_{\nu_{e}}$ neutrino mass scale and the sum of the neutrino 
  masses, we have
  \begin{align}
   m_{\nu_{e}} \approx m_{0} \left[ 1 + \frac{2}{3} R_{2} + \frac{1}{2} 
   \left( 1 + 4R_{3} \right) \sin^{2}{\theta_{13}} \right], \qquad 
   \sum_{i} \vert m_{\nu_{i}} \vert \approx 3 m_{0} \left[ 1 + \frac{2}{3} 
   \left( R_{2} + R_{3} \right)\right].
  \end{align}	
  Therefore, we expect that $m_{\nu_{e}} \geq 0.102~eV$ and 
  $\sum_{i} \vert m_{\nu_{i}} \vert \geq 0.312~eV$ with $m_{0}\gtrsim 0.1~eV$. 
  In this ordering, the bound in the neutrino mass scale may be reached if the common 
  neutrino mass is large, at the same time, the sum of the neutrino masses is far away 
  from the bound given by {\bf Planck} collaboration.
  \begin{figure}[ht]\centering
   \includegraphics[scale=0.6]{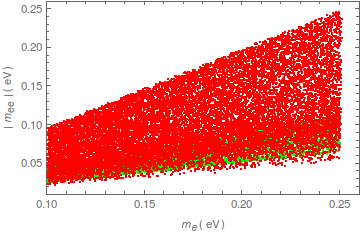}\hspace{0.5cm}\includegraphics[scale=0.6]{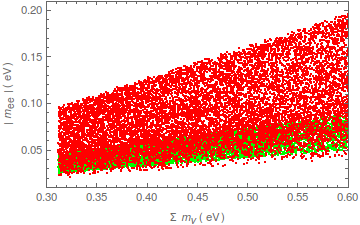}
   \caption{Predicted region for $\vert m_{ee}\vert$ versus the neutrino mass scale and the 
   sum of neutrino masses.}\label{mee4}
  \end{figure}

  Similarity to the inverted hierarchy, we add two plots where the atmospheric angle and the 
  effective neutrino mass are displayed against the relative Majorana phases. As we can see, 
  in the first panel, for the atmospheric angle the allowed region has been increased by the 
  presence of CP non trivial values in the Majorana phases; in this framework, two notable 
  regions seem remarkable: $a)$~$\alpha=\pi$ and $\beta=0,2\pi$; this values correspond to 
  the {\bf Case C}; 
  $b)$~$\alpha,\beta=\pi$, this CP parties correspond to the {\bf Case D}. These 
  two regions can be distinguished better in the second plot where the allowed values of the 
  effective neutrino mass is shown, in this plot the {\bf Case C} seems to be relevant in the 
  presence of CP non trivial values; in the case of CP parities, this case was disfavored by 
  predicting large values for the atmospheric angle.
  \begin{figure}[ht]\centering
\includegraphics[scale=0.6]{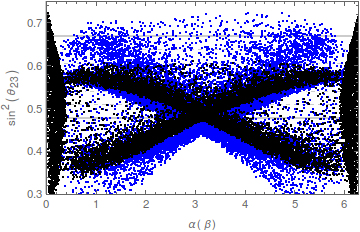}\hspace{0.5cm}\includegraphics[scale=0.6]{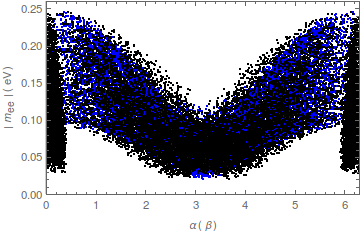}
\caption{Blue and black points stand for $\alpha$ and $\beta$ phases, respectively.}\label{mee6}
\end{figure}

As we can notice, the presence of CP non trivial values for the Majorana phases is a subtle issue since these change the allowed regions for the observables. We ought to comment that an analytical study on the CP non trivial phases should be carried out to verify the naive numerical results. For the moment, this task is out of the scope of this work.
\end{enumerate}
\section{Outlook and remarks}
We have constructed a $\mathbf{Q}_{6}$ flavored supersymmetric model with non-minimal Higgs sector. 
In the lepton sector, the $\lra$ permutation symmetry is broken only by one perturbative parameter $\epsilon$ which is directly proportional to the inequality $m_{e\tau}\neq m_{e\mu}$. 
This symmetry breaking deviates the reactor and atmospheric mixing angles from $0^{\circ}$ and $45^{\circ}$, respectively. Such deviations may be sizeable in the presence of CP parities in the Majorana phases, also this particular framework allows us to constrain, in an analytical way, the lightest neutrino mass and the $\epsilon$ parameter that accommodate the deviated mixing angles.

The model predicts that the inverted hierarchy is lesser favored than the degenerate ordering. For this latter mass ordering, the numerical values 
obtained for the reactor and atmospheric mixing angle are in very good agreement with the 
current experimental data on neutrino oscillations. Remarkably, the common neutrino mass lies on the favorable region for KamLAND-Zen 
collaboration and the atmospheric angle lies on the upper octant ($\theta_{23}> 45^{\circ}$).

Additionally, the effective neutrino mass decay, the neutrino mass scale and the sum of the neutrino mass were calculated in the framework of CP parities in the Majorana phases. In this particular case, the allowed regions are consistent with the analytical study, however, these regions are enhanced by the contribution of two relative Majorana phases in the neutrino masses. For the effective neutrino mass decay, the allowed values can be reached by the future experiments in this direction.


%
%
%

\section*{Acknowledgements}
This work was partially supported by the Mexican grants 237004, PAPIIT IN111115 and 
Conacyt  132059. 
JCGI thanks CINVESTAV for the warm hospitality and Red de Altas Energ\'{\i}as-CONACYT for the financial support. 
FGC acknowledges the financial support from {\it CONACYT} and {\it PRODEP} under Grant 
No.~511-6/17-8017.

\appendix
\section{$\mathbf{Q}_{6}$ Flavor Symmetry}

The $\mathbf{Q}_{6}$ group has twelve elements which are contained in six conjugacy 
classes, therefore, it contains six irreducible representations. We will use 
the notation given in~\cite{Kajiyama:2007pr}, there are various notations and 
extensive studies for this group, see for 
example~\cite{Kubo:2012ty,Ishimori:2010au}. The $\mathbf{Q}_{6}$ family symmetry has 
${\bf 2}$ two-dimensional irreducible representations denoted by 
${\bf 2}_{1}$ and ${\bf 2}_{2}$, $4$ one-dimensional ones which are denoted 
by ${\bf 1}_{+,0}$, ${\bf 1}_{+,2}$, ${\bf 1}_{-,1}$ and ${\bf 1}_{-,3}$. As 
it is well known, ${\bf 2}_{1}$ is a pseudo real and ${\bf 2}_{2}$ is a real 
representation. In addition, for ${\bf 1}_{\pm,n}$ we have that $n=0,1,2,3$ 
is the factor $\exp\left(in\pi/2\right)$ that appears in the matrix given by 
${\bf B}$. The $\pm$ stands for the change of sign under the transformation 
given by the ${\bf A}$ matrix. So that the first two one-dimensional 
representations are real and the two latter ones are complex conjugate to 
each other.
\begin{equation}
	\mathbf{Q}_{6} = \lbrace {\bf 1}, {\bf A}, {\bf A}^{2}, {\bf A}^{3}, {\bf A}^{4}, 
	{\bf A}^{5}, {\bf B}, {\bf A}{\bf B}, {\bf A}^{2}{\bf B}, {\bf A}^{3} 
	{\bf B}, {\bf A}^{4} {\bf B}, {\bf A}^{5} {\bf B} \rbrace,
\end{equation}
where the ${\bf A}$ and ${\bf B}$ are two-dimensional matrices whose explicit 
forms are given by
\begin{equation}
	{\bf A} = 
	\begin{pmatrix} 
		\cos\left(\pi/3 \right) & \sin\left(\pi/3 \right) \\ 
		-\sin\left(\pi/3 \right) & \cos\left(\pi/3 \right)
	\end{pmatrix}
	\quad \textrm{and} \quad
	{\bf B} = 
	\begin{pmatrix}
		i & 0 \\ 
		0 & -i
	\end{pmatrix} .
\end{equation}
Let us write the multiplication rules among the six irreducible 
representations which will be useful to build a phenomenological model: 
\begin{align}\label{q6rul}
		{\bf 1}_{+,2}\otimes{\bf 1}_{+,2} & =  {\bf 1}_{+,0},\quad {\bf 1}_{-,3} \otimes {\bf 1}_{-,3} 
		= {\bf 1}_{+,2},\quad {\bf 1}_{-,1} \otimes {\bf 1}_{-,1} = {\bf 1}_{+,2},\quad {\bf 1}_{-,1} 
		\otimes {\bf 1}_{-,3} = {\bf 1}_{+,0},\nn\\
		{\bf 1}_{+,2} \otimes {\bf 1}_{-,1} &=  {\bf 1}_{-,3},\quad {\bf 1}_{+,2} \otimes {\bf 1}_{-,3} 
		= {\bf 1}_{-,1},\quad {\bf 2}_{1} \otimes  {\bf 1}_{+,2} = {\bf 2}_{1},\quad {\bf 2}_{1} \otimes 
		{\bf 1}_{-,3} = {\bf 2}_{2},\nn\\
		{\bf 2}_{1} \otimes {\bf 1}_{-,1} &=  {\bf 2}_{2},\quad {\bf 2}_{2} \otimes {\bf 1}_{+,2} = {\bf 2}_{2},\quad
		{\bf 2}_{2} \otimes {\bf 1}_{-,3} = {\bf 2}_{1},\quad {\bf 2}_{2} \otimes {\bf 1}_{-,1} = {\bf 2}_{1}; \nn\\
		\overbrace{\begin{pmatrix}
				x_{1} \\ 
				x_{2}
			\end{pmatrix}}^{{\bf 2}_{1}}
			\, \otimes \, 
			\overbrace{\begin{pmatrix}
					y_{1} \\ 
					y_{2}
				\end{pmatrix}}^{{\bf 2}_{1}} & =  
				\overbrace{\left(x_{1}y_{2}-x_{2}y_{1}\right)}^{{\bf 1}_{+,0}} \quad +\quad \overbrace{\left(x_{1}y_{1}+x_{2}y_{2}\right)}^{{\bf 1}_{+,2}} \quad + \quad
				\overbrace{\begin{pmatrix}
						-x_{1}y_{2}-x_{2}y_{1} \\ 
						x_{1}y_{1}-x_{2}y_{2}
					\end{pmatrix}}^{{\bf 2}_{2}} \nn\\
					\overbrace{\begin{pmatrix}
							a_{1} \\ 
							a_{2}
						\end{pmatrix}}^{{\bf 2}_{2}} \otimes 
						\overbrace{\begin{pmatrix}
								b_{1} \\ 
								b_{2}
							\end{pmatrix}}^{{\bf 2}_{2}}&=\overbrace{\left(a_{1}b_{1}+a_{2}b_{2}\right)}^{{\bf 1}_{+,0}}\quad + \quad
							\overbrace{\left(a_{1}b_{2}-  a_{2}b_{1}\right)}^{{\bf 1}_{+,2}} \quad + \quad
							\overbrace{\begin{pmatrix}
									-a_{1}b_{1}+a_{2}b_{2} \\ 
									a_{1}b_{2}+a_{2}b_{1}
								\end{pmatrix}}^{{\bf 2}_{2}} \nn\\
								\overbrace{\begin{pmatrix}
										x_{1} \\ 
										x_{2}
									\end{pmatrix}}^{{\bf 2}_{1}} \otimes 
									\overbrace{\begin{pmatrix}
											a_{1} \\ 
											a_{2}
										\end{pmatrix}}^{{\bf 2}_{2}} &=\overbrace{\left(x_{1}a_{2}+x_{2}a_{1}\right)}^{{\bf 1}_{-,3}}\quad + \quad  
										\overbrace{\left(x_{1}a_{1}-x_{2}a_{2}\right)}^{{\bf 1}_{-,1}} \quad + \quad
										\overbrace{\begin{pmatrix}
												x_{1}a_{1}+x_{2}a_{2} \\ 
												x_{1}a_{2}-x_{2}a_{1}		\end{pmatrix}}^{{\bf 2}_{1}},
\end{align}
\bibliographystyle{bib_style_T1}
\bibliography{references.bib}

\end{document}